\documentclass[twocolumn]{aastex63}
\usepackage{natbib}
\usepackage{color}
\usepackage{mathtools}
\usepackage{hyperref} 
\hypersetup{colorlinks=true,citecolor=blue}

\graphicspath{{./Figures/}}  

\def    \apjl  		{\rm {ApJL}}
\def    \apj  		{\rm {ApJ}}
\def    \mnras  	{\rm {MNRAS}}
\def    \araa  		{\rm {ARA\&A}}
\def    \aa  		{\rm {A\&A}}
\def    \apjl  		{\rm {ApJL}}

\def	\cm		{\,{\rm {cm}}}
\def	\K		{\,{\rm K}}
\def	\g		{\,{\rm {g}}}
\def	\mum	{\,{\mu \rm{m}}}

\def \bea {\begin{eqnarray}}
\def \ena {\end{eqnarray}}                  

\def \bea {\begin{eqnarray}}
\def \ena {\end{eqnarray}}

\def	\B	{{\rm B}}

\def	\C	{{\rm C}}

\def	\cm	{\,{\rm cm}}
\def	\d	{{\rm d}}

\def	\D	{{\rm D}}
\def	\eff	{{\rm eff}}

\def	\erg	{\,{\rm erg}}

\def	\g	{\,{\rm g}}
\def	\gas	{\,{\rm gas}}
\def	\km	{\,{\rm km}}
\def	\G	{{\rm G}}

\def	\H	{{\rm H}}

\def	\N	{{\rm N}}

\def	\s	{\,{\rm s}}

\def	\AU	{\,{\rm au}}

\def	\H	{{\rm H}}
\def	\O	{{\rm O}}

\def	\V	{{\rm V}}

\def	\rad	{\,{\rm rad}}
\def	\yr	    {\,{\rm yr}}

\def    \gas     	{{\rm gas}}

\begin{document}
\shorttitle{Disruption of cometary dust and ice}
\shortauthors{Hoang and Tung}
\title{Evolution of Dust and Water Ice in Cometary Comae by Radiative Torques}

\author{Thiem Hoang}
\affil{Korea Astronomy and Space Science Institute, Daejeon 34055, Republic of Korea}
\affil{University of Science and Technology, Korea, (UST), 217 Gajeong-ro Yuseong-gu, Daejeon 34113, Republic of Korea}

\author{Ngo-Duy Tung}
\affil{University of Science and Technology of Hanoi, VAST, 18 Hoang Quoc Viet, Hanoi, Vietnam}

\begin{abstract}
Comets provide unique information about the physical and chemical properties of the environment in which the Solar system was formed. Understanding how cometary dust and ice evolve under the effect of sunlight is essential for constraining nuclear structure and triggering mechanism of comet activity. In this paper, we first study rotational disruption of dust grains lifted by outgassing from comet nuclei by radiative torques (RATs). We find that composite grains could be rapidly disrupted into small fragments by the Radiative Torque Disruption (RATD) mechanism. We then study rotational desorption of ice grains by RATs and find that icy grains could be desorbed from large heliocentric distances, beyond the sublimation radius of water at $R_{\rm sub}(\H_{2}\O)\sim 3\AU$. We also calculate the production rate of water vapor versus the heliocentric distance of comets due to rotational desorption. Our results could explain the variation of dust properties and the presence of small grains frequently observed from cometary comae. Finally, we suggest that the activity of distant comets could be triggered by rotational disruption of grains and desorption of water ice grains at large heliocentric distances.
 
\end{abstract}

\section{Introduction}\label{sec:intro}
Comets provide essential information about the physical and chemical properties of the environment in which the Solar system was formed. In general, a comet seen in the night sky includes a small nucleus, an extended coma, and long tails (ion and dust). Comet nuclei are made of dust, water ice, and frozen compounds, which are implied from the ``dirty snowball'' model \citep{1950ApJ...111..375W} or ``icy dirtball'' model \citep{1989ESASP.302...39K}. Understanding how comet activity is triggered is a longstanding problem in cometary science (see \citealt{2018SSRv..214...64L} and \citealt{2020SSRv..216...14K} for recent reviews). 

The current paradigm for the comet activity is based on thermal sublimation of frozen compounds present in the comet nucleus. For most of their orbits, comets are far away from the Sun and stay inactive. When they come closer, highly volatile ices such as CO, CO$_2$, CH$_{4}$, NH$_{3}$, first evaporate due to solar heating. Outflowing gas thus drags dust and water ice grains off the nucleus, triggering the activity of comets. Such dust and ice grains scatter sunlight, producing an extended glowing region known as the cometary coma. 

The properties of dust (size, shape, and composition) and ice in the coma are crucially important for understanding and interpreting observational data. Dust grains in comet nuclei are widely thought of as an agglomerate of small monomers \citep{2019A&A...630A..24G}. In the coma, dust properties are expected to changed when embedded volatiles sublimate, whereas water ice grains are expected to be long-lived at large heliocentric distances, beyond the sublimation zone of heliocentric distance of $R_{\rm sub}(\H_{2}\O)\sim 3\AU$ (\citealt{1981PThPS..70...35H}; \citealt{1985A&A...142...31Y}) due to low temperatures (see \citealt{1976ARA&A..14..143W}; \citealt{AHearn:2011ch}; \citealt{Mumma:2011jn} for reviews), or even close to the Sun in the case of pure ice particles such as those observed in comet 103P/Hartley 2 by EPOXI at $1.064 \AU$ \citep{2011Sci...332.1396A}.

Optical-near infrared (NIR) polarimetric observations of scattered sunlight are useful to constrain dust properties from cometary comae. Various observations suggest that the size distribution of cometary dust grains varies over time (see \citealt{2004come.book..565F} and \citealt{2004come.book..577K}). 
\cite{Gicquel:2012fk} also found evidence of fragmentation of large grains into smaller ones for this comet. The authors found that small grains of mass $m<10^{-14}\g$ (i.e., radius $a<0.15\mum$) are much more abundant than predicted by theoretical models. Observations from comet Hale-Bopp by \cite{2000Icar..143..338J} also suggest grains either compact of size $a<0.5\mum$ or aggregates of $a<0.5\mum$. Optical-NIR polarimetric observations by \cite{Jones:2008jp} reveal the presence of small grains in the cometary coma of COMET 73P/SCHWASSMANN-WACHMANN 3 (see also \citealt{Rosenbush:2017}; \citealt{Kiselev:2020}). Furthermore, small grains are detected in this comet through the 10 $\mum$ silicate emission feature \citep{2001ApJ...549..635M}.
Large grains which are lifted off cometary nuclei are believed to be transformed into smaller ones by means of thermal sublimation/desorption as a volatile ``glue,'' which keeps together the grains in aggregates, evaporates. However, the feasibility and efficiency of such a mechanism remains unclear.

In this paper, we will study the evolution of dust grains from cometary comae using the new effect of RAdiative Torque Disruption (RATD) discovered by \cite{Hoang:2019da} (see also \citealt{2019ApJ...876...13H}). The RATD mechanism is based on the fact that dust grains of irregular shapes exposed to anisotropic radiation field experience Radiative Torques (RATs; \citealt{Dolginov:1976p2480}; \citealt{1996ApJ...470..551D}; \citealt{2007MNRAS.378..910L}; \citealt{Hoang:2008gb}). RATs can spin up the grain to suprathermal rotation (\citealt{1996ApJ...470..551D}; \citealt{2004ApJ...614..781A}) such that the resulting centrifugal stress can exceed the maximum tensile strength of the grain material, which breaks the grain into small fragments (\citealt{Hoang:2019da}). Due to its proximity to the radiation source and abundant data from in-situ measurements by spacecraft and remote observations, cometary comae provide a unique test for the RATD mechanism. Very recently, \cite{Herranen:2020im} studied rotational disruption of fluffy dust grains in comets using radiative torques obtained from numerical calculations and found that rotational disruption is efficient. However, the author did not study the disruption of dust across the coma and disregarded ice grains.

Water ice is an important component of cometary nuclei, which is first proposed seven decades ago by ``dirty snowball model'' by \cite{1950ApJ...111..375W} where the nucleus is proposed as a conglomerate of dust and ices. However, recently, a ``icy dirtballs'' model is introduced based on observations (see \citealt{2020SSRv..216...14K} for a review). The possible detection of icy grains in cometary nuclei is studied in \cite{1981Icar...47..342H} based on thermal sublimation. 
The existence of icy grains in comets is now established through direct detection by instruments onboard spacecraft (Deep Impact, \citealt{2006A&A...448L..53S}; EPOXI, \citealt{AHearn:2011dn}; Rosetta, \citealt{Schulz:2015il}) or spectroscopic observations of water ice absorption features (e.g., \citealt{2009AJ....137.4538Y}). In particular, \cite{Protopapa:2014} report that icy particles in comet Hartley 2 were aggregates of $\sim 1\mum$ pure water-ice constituent grains.

Understanding how and where water ice is transformed into water vapor is essential for accurate determination of water ice content. The current paradigm is that the activity of comets is triggered by thermal sublimation of water ice. However, as shown in \cite{2020ApJ...891...38H} and \cite{2019ApJ...885..125H}, ice mantles could be desorbed by rotational desorption at lower temperatures than classical sublimation. Thus, the rotational desorption would dramatically affect the water production rate and accurate determination of water content in comet nuclei from observations. We will quantify the effect of rotational desorption of ice grains in comae.

The structure of the paper is as follows. In Section \ref{sec:model}, we show the comet model adopted in our paper. In Section \ref{sec:dust}, we review the disruption mechanism of dust grains and present numerical results. Section \ref{sec:ice} is devoted to studying rotational desorption of water ice grains. An extended discussion of our results is presented in Section \ref{sec:discuss}. A summary of our main results is presented in Section \ref{sec:summary}.

\section{Physical model of a cometary coma}\label{sec:model}
The cometary coma is assumed to be spherical in which gas and dust are being produced continuously from the nucleus due to heating by solar radiation. Special comet activities like jets, fans, etc. are neglected in the scope of this paper. Highly volatile ice, such as CO and CO$_{2}$, is expected to evaporate first and lift dust and water ice grains off the nucleus. Since the nucleus presumes to be heated symmetrically by sunlight because its rotation period ($\sim 10^{4}\s$) is much shorter than the orbital period ($\sim 10^{6}\s$), gas and dust are expanding symmetrically in the radial direction.

Let $Q_{\gas}$ be the rate of mass production by the cometary nucleus and $v_{\gas}$ be the expansion velocity of gas. The gas mass density at distance $r$ from the nucleus (i.e., cometocentric distance) can be described by the Haser model (\citealt{Haser:1957uea}; see also \citealt{1985AJ.....90.2609C}):
\bea
\rho_{\gas}=\frac{Q_{\gas}}{4\pi v_{\gas} 
r^{2}}\exp\left(-\frac{r}{L_{g}}\right),\label{eq:ngas_coma}
\ena
where $dM=\rho_{\gas}4\pi r^{2}v_{\gas}dt=Q_{\gas}dt$ is the mass produced during the time interval $dt$, $L_{g}$ is the ionization length scale, which is between $1-2\times 10^{6}\km$ (see \citealt{1991JGR....96.7731L}). Above, the exponential term describes the decay of gas, and the subdominant effect of solar radiative pressure on the expanding gas is disregarded. For a coma with the radius $r\ll L_{g}$, we can ignore the exponential term in Equation (\ref{eq:ngas_coma}). Physical parameters for a coma are listed in Table \ref{tab:comet}. 

The gas production rate increases with decreasing distance $R$ from the Sun (hereafter heliocentric distance) due to the dependence of radiation flux as $1/R^{2}$. In general, $Q_{\rm gas}$ increases with decreasing the heliocentric distance $R$ due to solar radiation as 
\bea
Q_{\gas}=Q_{0}\left(\frac{R}{1\AU}\right)^{-\alpha},\label{eq:Qgas}
\ena
where $Q_{0}$ is the gas mass rate at 1 AU, and $\alpha\sim 2-4$ (see \citealt{2001MNRAS.326..852S}). We assume a set of values of $Q_{0}=2.5\times 10^{5}\g\cm^{-3}$ and $\alpha = 3.7$ for the 1961 apparition of the Comet 2P/Encke \citep{2001MNRAS.326..852S}.

The decrease of the radiation flux with increasing heliocentric distance also results in the variation of the gas expansion velocity with $R$, which can be estimated by a power law \citep{Delsemme1982} as
\bea
v_{\gas}\simeq 0.58\left(\frac{R}{1 \AU}\right)^{-0.5} \km\s^{-1},\label{eq:vgas}
\ena
and the variation of $v_{\rm gas}$ with the cometocentric distance is ignored for simplicity.

The number density of nucleon in the cometary coma is then given by 
\bea
n_{\H}=\frac{\rho_{\rm gas}}{\mu m_{\H}}=n_{0}\left(\frac{r}{r_{n}}\right)^{-2}\exp\left(-\frac{r}{L_{g}}\right),\label{eq:ngas}
\ena
where $n_{0}$ is the density at cometocentric distance $r_{n}$, $\mu$ is the mean molecular weight with $\mu=1$ for purely hydrogen gas and $\mu=1.4$ for gas of $10\%$ He, and $\mu=28$ for purely CO gas.

\begin{table}
\centering
\caption{Model parameters for a cometary coma}\label{tab:comet}
\begin{tabular}{l l}\hline\hline\\
Parameters & Values \cr
\hline\\
{Radius of nucleus, $r_{n}$}& {$1\km$}\cr
{Star temperature}& {$T_{\eff}=5800\K$}\cr
{Star luminosity}& {$L=L_{\odot}$}\cr
{Star radius}& {$R=R_{\odot}$}\cr
{Gas density}& {$n=n_{0}\left(r_{n}/r\right)^{2}$} $^{\rm a}$\cr
{Gas temperature}& {$T_{\gas}=300\K$}\cr
{Expansion velocity}& {$v_{\gas}= 0.58(R/1 \AU)^{-0.5} \km\s^{-1}$}\cr
\cr
\hline\\
\multicolumn{2}{l}{$^{\rm a}$ Here $n_{0}=4\times 10^{12} \cm^{-3}$ at $r_{n}=1$ km for $Q_{\rm gas}=9\times 10^{4}\g\s^{-1}$.}
\end{tabular}
\end{table}


\section{Rotational disruption of cometary dust}\label{sec:dust}
We first study disruption of aggregate dust grains due to radiative torques. 

\subsection{Rotational disruption mechanism}
The basic idea of the rotational disruption mechanism is as follows. A spherical dust grain of mass density $\rho$ rotating at angular velocity $\omega$ develops a centrifugal stress due to centrifugal force, which scales as $S=\rho a^{2} \omega^{2}/4$ (\citealt{Hoang:2019da}). When the rotation rate increases to a critical limit such that the tensile stress induced by centrifugal force exceeds the maximum tensile stress, the so-called tensile strength of the material, the grain is disrupted instantaneously. The critical angular velocity for the disruption is given by
\bea
\omega_{\rm cri}&=&\frac{2}{a}\left(\frac{S_{\rm max}}{\rho} \right)^{1/2}\nonumber\\
&\simeq& \left(\frac{3.6\times10^{9}}{a_{-5}}\right)\hat{\rho}^{-1/2}S_{\rm max,9}^{1/2}~\rm rad/s,~~~~\label{eq:omega_cri}
\ena
where $a_{-5}=a/(10^{-5}\cm)$, $\hat{\rho}=\rho/(3\g\cm^{-3})$ with $\rho$ being the dust mass density, and $S_{\rm max}$ is the tensile strength of dust material and $S_{\rm max,9}=S_{\rm max}/(10^{9}\erg\cm^{-3})$ is the tensile strength in units of $10^{10}\erg\cm^{-3}$. The exact value of $S_{\max}$ depends on the dust grain composition and structure. Compact grains can have higher $S_{\max}$ than porous/composite grains. Ideal material without impurity, such as diamond, can have $S_{\max}\ge 10^{11}\erg\cm^{-3}$ (see \citealt{Hoang:2019da} for more details). 

\subsection{Rotation rate of irregular grains spun-up by radiative torques}
Grains subject to the anisotropic radiation field experience RATs which act to spin-up the grains to suprathermal rotation (\citealt{Dolginov:1976p2480}; \citealt{1996ApJ...470..551D}; \citealt{2007MNRAS.378..910L}; \citealt{Hoang:2008gb}). To describe the strength of a radiation field, let define $U=u_{\rm rad}/u_{\rm ISRF}$ with 
$u_{\rm ISRF}=8.64\times 10^{-13}\erg\cm^{-3}$ being the energy density of the average interstellar radiation field (ISRF) in the solar neighborhood as given by \cite{1983A&A...128..212M}. Thus, the typical value for the ISRF is $U=1$. We consider comets approaching the Sun with $L_{\star}=L_{\odot}, R_{\star}=R_{\odot},~T_{\star}=5800\K,~M_{\star}=M_{\odot}$.

Let $u_{\lambda}$ be the spectral energy density of radiation field at wavelength $\lambda$. The radiation energy density at heliocentric distance $R$ is given by
\bea
u_{\rm rad}=\frac{L_{\star}}{4\pi R^{2}c}\simeq 4.5\times 10^{-7}\left(\frac{L_{\star}}{L_{\odot}}\right)\left(\frac{R}{1\AU}\right)^{-2} \erg\cm^{-3}.
\ena
which corresponds to 
\bea
U=5.2\times 10^{7}\left(\frac{R}{1\AU}\right)^{-2}.
\ena

The mean wavelength of the radiation field is 
\bea
\bar{\lambda}=\frac{\int u_{\lambda}\lambda d\lambda}{\int u_{\lambda}d\lambda}
\ena
which yields $\bar{\lambda}\sim 0.91\mum$ for the solar-type star.

In the plasma, grain rotation experiences damping due to collisions with gas species. The well-known damping process for a rotating grain is sticking collisions with gas atoms, followed by thermal evaporation. Thus, for a gas with He of $10\%$ abundance, the characteristic damping time is
\bea
\tau_{\gas}&=&\frac{3}{4\sqrt{\pi}}\frac{I}{1.2n_{\rm H}m_{\rm H}
v_{\rm th}a^{4}}\nonumber\\
&\simeq& 2200\left(\frac{\hat{\rho}a_{-5}}{n_{8}T_{2}^{1/2}}\right)~{\rm days},~~
\ena
where $v_{\rm th}=\left(2k_{\B}T_{\rm gas}/m_{\rm H}\right)^{1/2}$ is the thermal velocity of a gas atom of mass $m_{\rm H}$ in a plasma with temperature $T_{\gas}$ and nucleon density $n_{\H}$, the spherical grains are assumed (\citealt{2009ApJ...695.1457H}; \citealt{1996ApJ...470..551D}). Above, $n_{8}=n_{\H}/(10^{8}\cm^{-3})$, $T_{2}=T_{\gas}/100\K$.

This time is equal to the time required for the grain to collide with an amount of gas of the grain mass.

IR photons emitted by the grain carry away part of the grain's angular momentum, resulting in the damping of the grain rotation. For strong radiation fields or not very small sizes, grains can achieve equilibrium temperature, such that the IR damping coefficient (see \citealt{1998ApJ...508..157D}) can be calculated as
\bea
F_{\rm IR}\simeq 5.5\times 10^{-3}\left(\frac{U_{7}^{2/3}}{n_{8}T_{2}^{1/2}}\right)a_{-5},\label{eq:FIR}
\ena 
which implies subdominance of the IR damping over the gas damping for $n_{\H}>10^{6}\cm^{-3}$ and $U<10^{7}$. 

Other rotational damping processes include plasma drag, ion collisions, and electric dipole emission. These processes are mostly important for PAHs and very small grains (\citealt{1998ApJ...508..157D}; \citealt{Hoang:2010jy}; \citealt{2011ApJ...741...87H}). Thus, the total rotational damping rate by gas collisions and IR emission can be written as
\bea
\tau_{\rm damp}^{-1}=\tau_{\gas}^{-1}(1+ F_{\rm IR}).\label{eq:taudamp}
\ena

For the radiation source with stable luminosity considered in this paper, radiative torques $\Gamma_{\rm RAT}$ is constant, and the grain velocity is steadily increased over time. The equilibrium rotation can be achieved at (see \citealt{2007MNRAS.378..910L}; \citealt{2009ApJ...695.1457H}; \citealt{2014MNRAS.438..680H}):
\bea
\omega_{\rm RAT}=\frac{\Gamma_{\rm RAT}\tau_{\rm damp}}{I},~~~~~\label{eq:omega_RAT0}
\ena
where $I=8\pi \rho a^{5}/15$ is the grain inertia moment.

Following \cite{Hoang:2019da} and \cite{2019ApJ...876...13H}, the rotation rate by RATs for an unidirectional radiation of $\gamma=1$ is given by
\bea
\omega_{\rm RAT}&\simeq &3.2\times 10^{8} a_{-5}^{0.7}\bar{\lambda}_{0.5}^{-1.7}\nonumber\\
&\times&\left(\frac{U_{7}}{n_{8}T_{2}^{1/2}}\right)\left(\frac{1}{1+F_{\rm IR}}\right)\rad\s^{-1},~~~\label{eq:omega_RAT1}
\ena
for grains with $a\lesssim a_{\rm trans}$, and
\bea
\omega_{\rm RAT}&\simeq &1.6\times 10^{9}\frac{1}{a_{-5}^{2}}\bar{\lambda}_{0.5}\nonumber\\
&&\times \left(\frac{U_{7}}{n_{8}T_{2}^{1/2}}\right)\left(\frac{1}{1+F_{\rm IR}}\right)\rad\s^{-1},~~~\label{eq:omega_RAT2}
\ena
for grains with $a> a_{\rm trans}$. Here, $\bar{\lambda}_{0.5}=\bar{\lambda}/(0.5\mum)$.

For convenience, let $a_{\rm trans}=\bar{\lambda}/1.8$ which denotes the grain size at which the RAT efficiency changes between the power law and flat stages (see e.g., \citealt{2007MNRAS.378..910L}; \citealt{{Hoang:2019da}}), and $\omega_{\rm RAT}$ changes from Equation (\ref{eq:omega_RAT1}) to (\ref{eq:omega_RAT2}).

\subsection{Grain disruption size}

By setting $\omega_{\rm RAT}=\omega_{\rm disr}$, one obtains the size where dust grains start to be disrupted by RATs for an arbitrary radiation field with $a\le a_{\rm trans}$:
\bea
\left(\frac{a_{\rm disr}}{0.1\mum}\right)^{1.7}&\simeq&0.11 \bar{\lambda}_
{0.5}^{1.7}S_{\max,7}^{1/2}\nonumber\\
&&\times (1+F_{\rm IR})\left(\frac{n_{8}T_{2}^{1/2}}{U_{7}}\right),~~~\label{eq:adisr_comp2}
\ena
which depends on the local gas density, temperature, and the grain tensile strength $S_{\max}$. Above, $S_{\rm max,7}=S_{\rm max}/(10^{-7}\erg\cm^{-3})$

Due to the decrease of the rotation rate for $a>a_{\rm trans}$, there exist a maximum size of grains that can still be disrupted by centrifugal stress (\citealt{2020ApJ...891...38H}):
\bea
a_{\rm disr,max}\simeq 5.0\gamma\bar{\lambda}_{0.5}\left(\frac{U_{7}}{n_{8}T_{2}^{1/2}}\right)\left(\frac{1}{1+F_{\rm IR}}\right)\hat{\rho}^{1/2}S_{\max,7}^{-1/2}~\mum.~~~~~\label{eq:adisr_up}
\ena 

In the presence of RATD, grains of sizes in the range $a_{\rm disr} \leq a < a_{\rm disr,max}$ will be disrupted. Equation (\ref{eq:adisr_comp2}) implies that grain disruption occurs as long as $n_{\H}T_{\gas}^{1/2}<U$. For comets at $R\sim 1-2 \AU$, $U\sim 5\times 10^{7}$. Therefore, grains in the $n_{\H}<5\times 10^{7}\cm^{-3}$ regions of the coma can be disrupted. Using Equation (\ref{eq:ngas_coma}) one can see that is equivalent to the regions of $r>250 \km$ for $a=0.1\mum$. Larger grains of $a\sim 0.5\mum$ can be disrupted at smaller distances from the nucleus.

\subsection{Disruption time and lifetime of grains}
In the absence of rotational damping, the characteristic timescale for rotational desorption can be estimated as:
\bea
t_{\rm disr,0}&=&\frac{I\omega_{\rm disr}}{dJ/dt}=\frac{I\omega_{\rm disr}}{\Gamma_{\rm RAT}}\nonumber\\
&\simeq& 2.1(\gamma U_{7})^{-1}\bar{\lambda}_{0.5}^{1.7}\hat{\rho}_{\rm ice}^{1/2}S_{\max,7}^{1/2}a_{-5}^{-0.7}{~\rm days}\label{eq:tdisr}
\ena
for $a_{\rm disr}<a \lesssim a_{\rm trans}$, and
\bea
t_{\rm disr,0}\simeq& 0.14(\gamma U_{7})^{-1}\bar{\lambda}_{0.5}^{-1}\hat{\rho}_{\rm ice}^{1/2}S_{\max,7}^{1/2}a_{-5}^{2}{~\rm days}
\ena
for $a_{\rm trans}<a<a_{\rm disr,max}$.

In the presence of rotational damping, the disruption timescale can be obtained by solving $\omega(t)=\omega_{\rm disr}$, which yields
\bea
t_{\rm disr}&=&-\tau_{\rm damp}\ln \left(1-\frac{\omega_{\rm disr}}{\omega_{\rm RAT}}\right)\nonumber\\
&=&-\tau_{\rm damp}\ln \left(1-\frac{t_{\rm disr,0}}{\tau_{\rm damp}}\right),\label{eq:tdisr_exact}
\ena
which is applicable for $a_{\rm disr,max}>a>a_{\rm disr}$. Note that $t_{\rm disr}\rightarrow \infty$ for $a=[a_{\rm disr}, a_{\rm disr,max}]$ because it takes $t\gg t_{\rm damp}$ to reach $\omega=\omega_{\rm RAT}$. One see that $t_{\rm disr}$ returns to $t_{\rm disr,0}$ when $t_{\rm disr,0}\ll \tau_{\rm damp}$ which is achieved in strong radiation fields.

\subsection{Numerical Results}
For composite grains considered here, we assume the typical radius of monomers $a_{p}=0.1\mum$ and the maximum grain size $a_{\max} = \bar{\lambda}/0.1 \sim 10\mum$. Our choice of $a_{\max}$ is based on the fact that rotational disruption is expected to drop for $a\gtrsim \bar{\lambda}/0.1 \sim 10\mum$ due to the decrease of RATs which originates from the canceling effect (\citealt{2007MNRAS.378..910L}) and the calculations of RATs for large grains of size $a > \bar{\lambda}/0.1$ are not yet available due to computing limitations (see e.g., \citealt{2019ApJ...878...96H}).

The tensile strength $S_{\max}$ of cometary dust can vary depending on composition, internal structure, and grain sizes (see, e.g., \citealt{2020MNRAS.496.1667K} for a review). \citet{2016P&SS..133...63H} estimated the low tensile strength of dust aggregates of sizes $a = 20-155\mum$ in Comet 67P/Churyumov-Gerasimenko, $S_{\max} \sim 10^{4}\erg\cm^{-3}$, whereas for Comet 17P/Holmes, $S_{\max} \sim 10^{4}-10^{6}\erg\cm^{-3}$ for $a=2-200\mum$ \citep{2010Icar..208..276R}. We note that the grain sizes considered in this paper ($a \leq 10\mum$) are much smaller than the dust aggregates studied from in situ measurements. Therefore, one can expect a larger value of tensile strength. To account for that, for large composite grains of $0.1\mum \leq a \leq 10\mum$, we let $S_{\max}$ vary between $10^{5}-10^{10}\erg\cm^{-3}$ (\citealt{2019ApJ...876...13H}). For small grains of $a<a_{p}$, which are expected to be compact, we fix the tensile strength to $S_{\rm max}\gtrsim 10^{9}\erg\cm^{-3}$ (\citealt{Hoang:2019da}).  

\begin{figure*}
\includegraphics[width=0.5\textwidth]{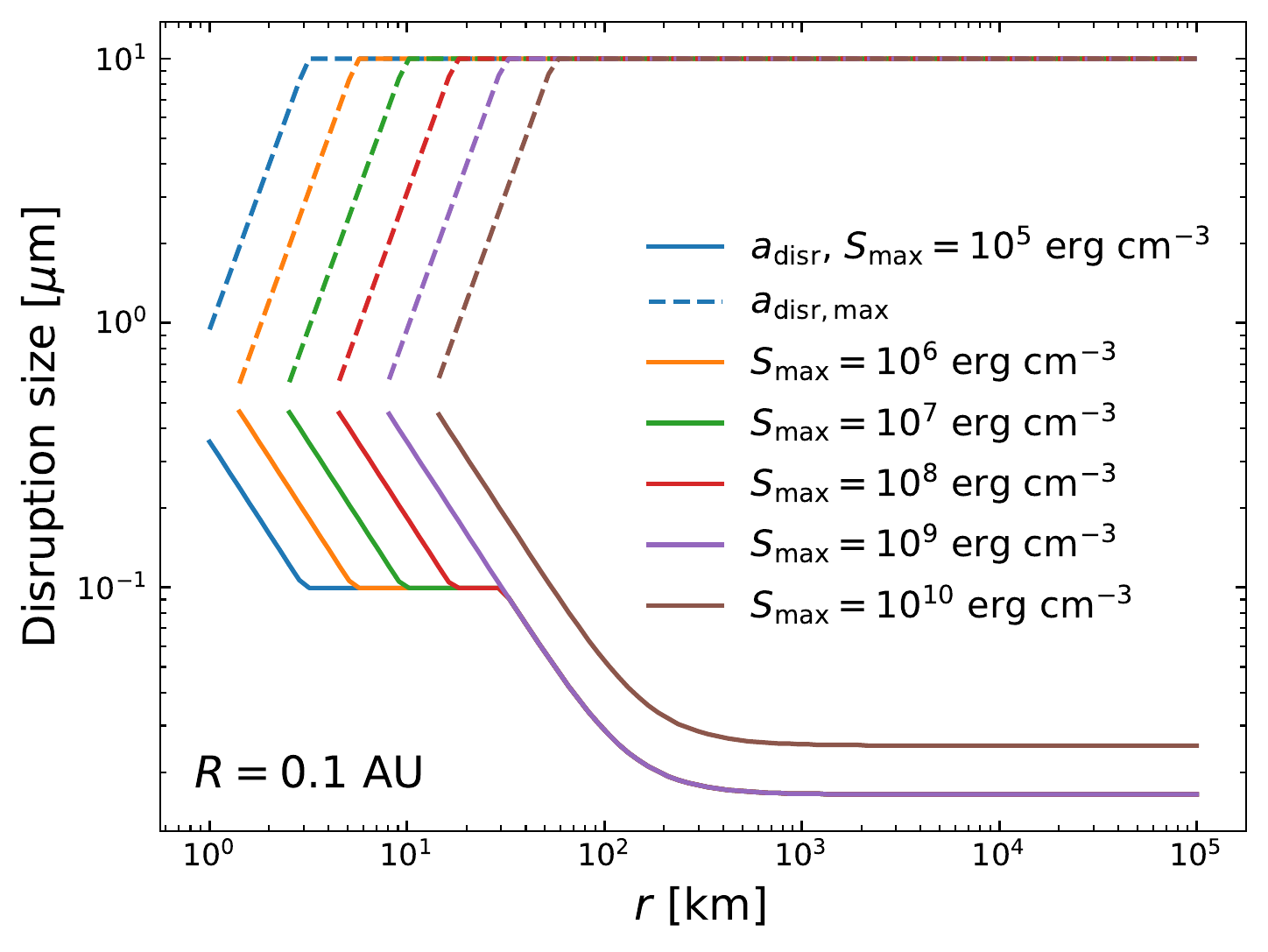}
\includegraphics[width=0.5\textwidth]{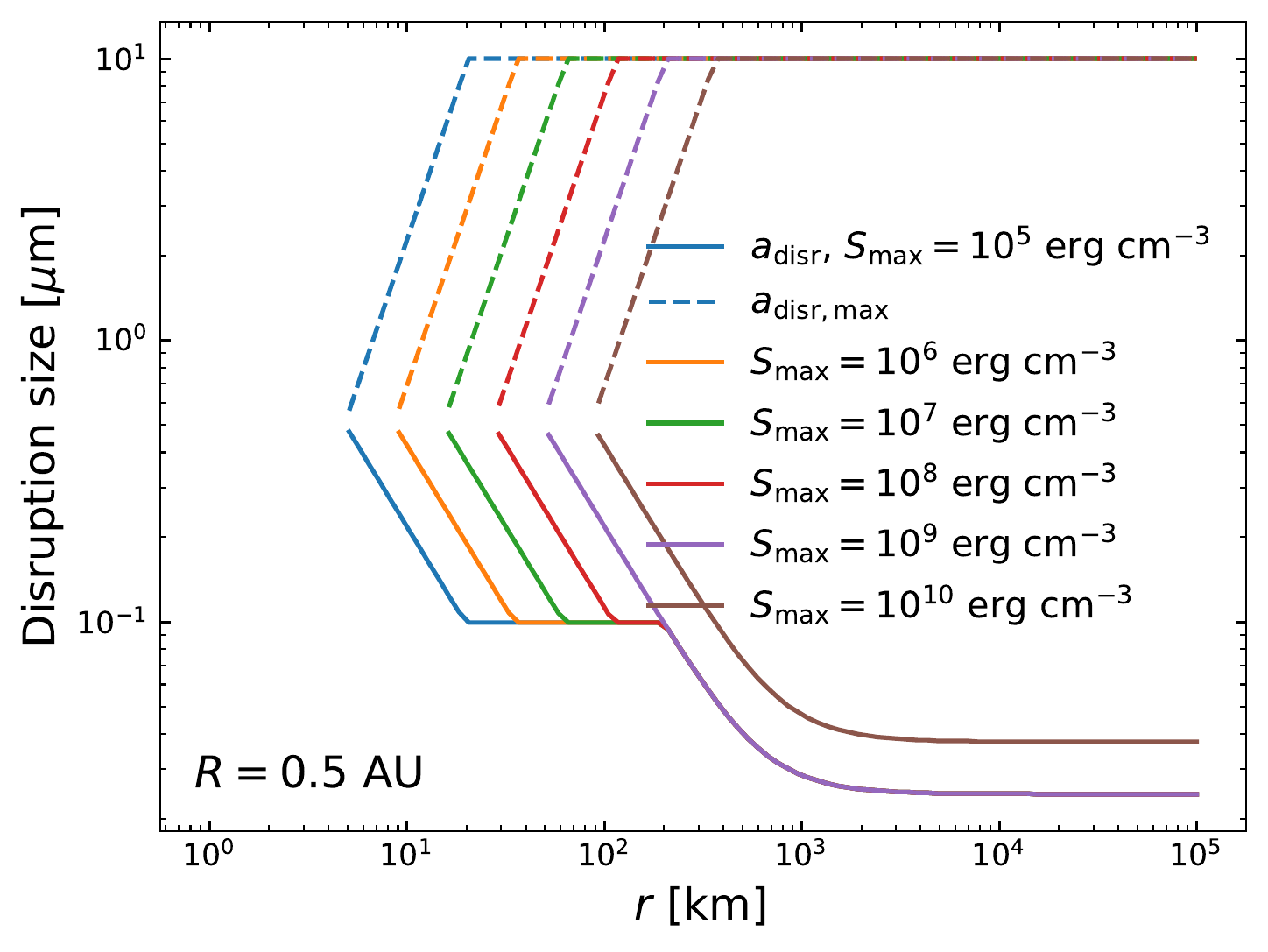}
\includegraphics[width=0.5\textwidth]{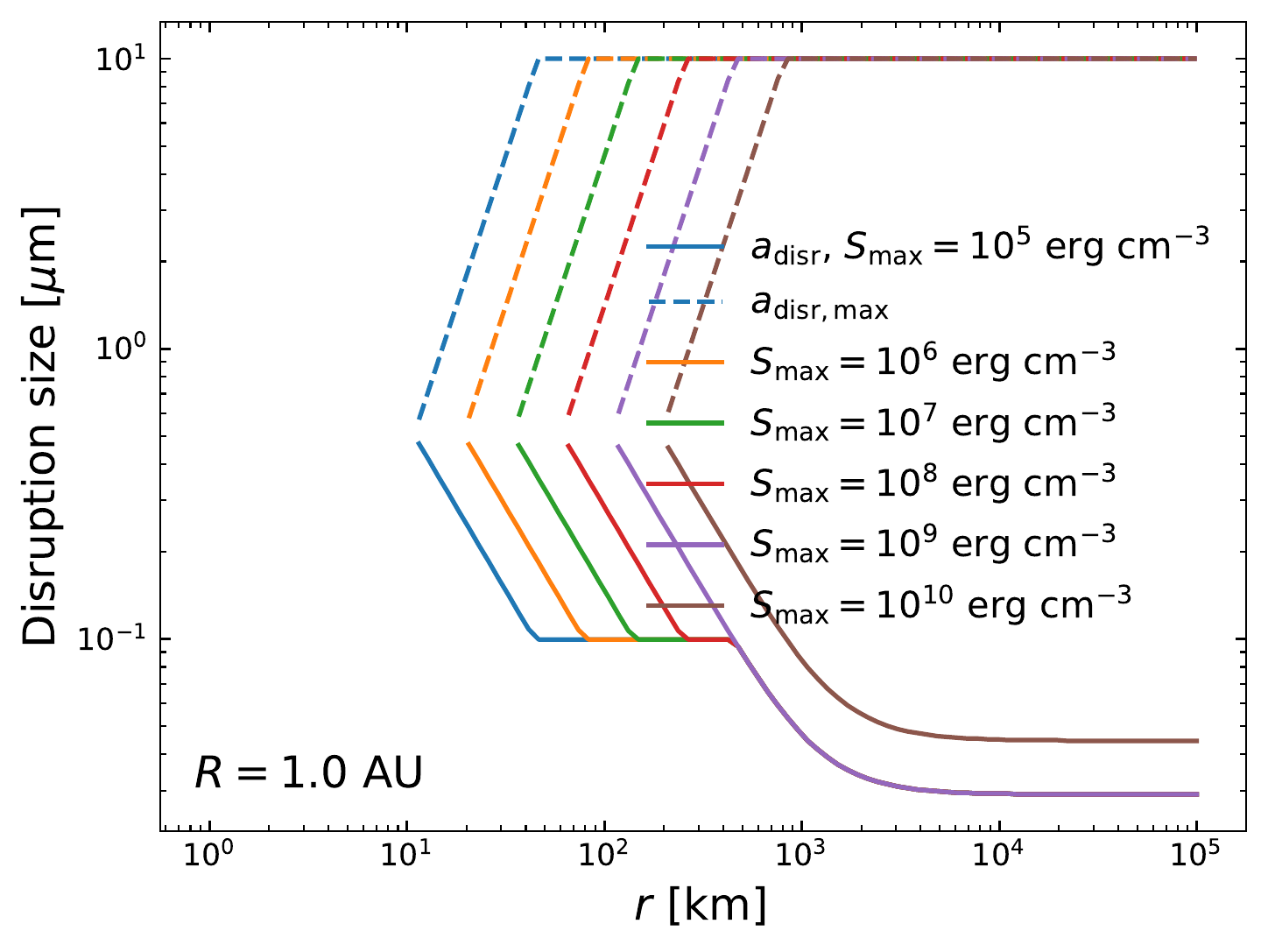}
\includegraphics[width=0.5\textwidth]{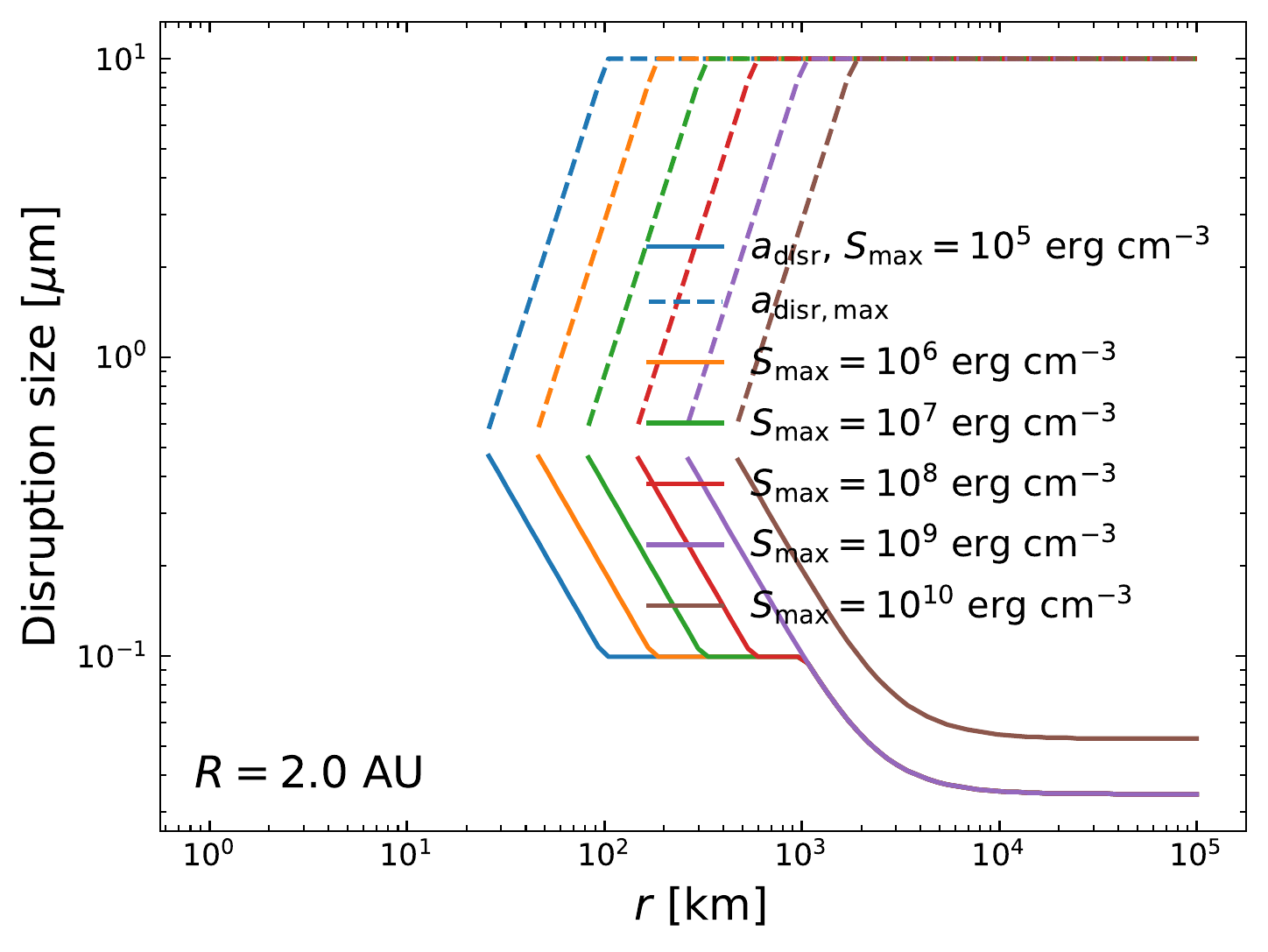}
\caption{Grain disruption size vs. cometocentric distance for different tensile strengths, $S_{\max}$, and constant $Q_{\rm gas}$. The comet is located at different heliocentric distances from $R=0.1-2$ AU. The disruption zone, located between the solid ($a_{\rm disr}$) and dashed ($a_{\rm disr,max}$) lines, is more extended for weaker grains or smaller $R$.}
\label{fig:adisr_Qconst}
\end{figure*}

Figure \ref{fig:adisr_Qconst} shows the grain disruption size as a function of the cometocentric distance $r$ for different $S_{\rm max}$ and heliocentric distances, $R$, assuming a typical constant value of $Q_{\rm gas} = 9 \times 10^{4} \g \s^{-1}$ \citep{{Rosenbush:2017}}. The disruption range is defined by the grain size between $a_{\rm disr}$ and $a_{\rm disr,max}$. For a given heliocentric distance, rotational disruption takes place in the outer region of the cometary coma where the gas density has considerably decreased. The grain disruption size, $a_{\rm disr}$, decreases rapidly with $r$, from $a_{\rm trans}$ = $0.51\mum$ to a minimum value $\sim a_{\rm disr}=a_{p}= 0.1\mum$ (see solid lines). At large $r$ where the gas density decreases to a small value, even compact grains of large tensile strengths $S_{\max}\gtrsim 10^{9}\erg\cm^{-3}$ could be disrupted. The maximum disruption size, $a_{\rm disr,max}$, increases rapidly with $r$ and reaches the threshold $a_{\rm max} = \bar{\lambda}/0.1\sim 10\mum$ (see dashed lines). Thus, grains of sizes between $a_{\rm disr}-a_{\rm disr,max}$ will be disrupted under the effect of RATD. Moreover, the disruption zone is more extended for grains having lower tensile strengths. At $R=2$ AU, grains can be disrupted only in the region of $r>10$ km, but at $R=0.1$ AU, grains can be disrupted at $r<10$ km due to larger solar radiation flux.

\begin{figure*}
\includegraphics[width=0.5\textwidth]{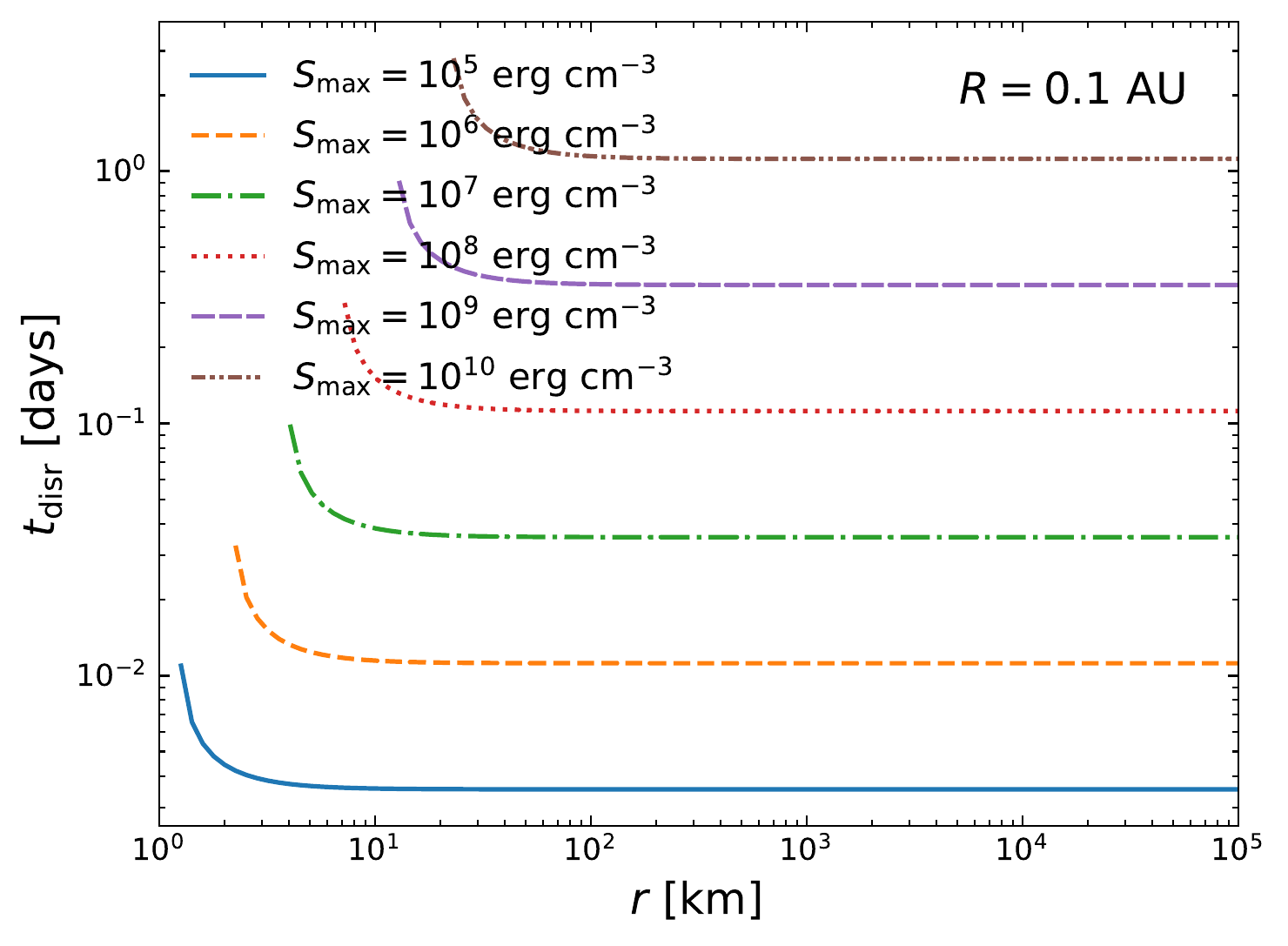}
\includegraphics[width=0.5\textwidth]{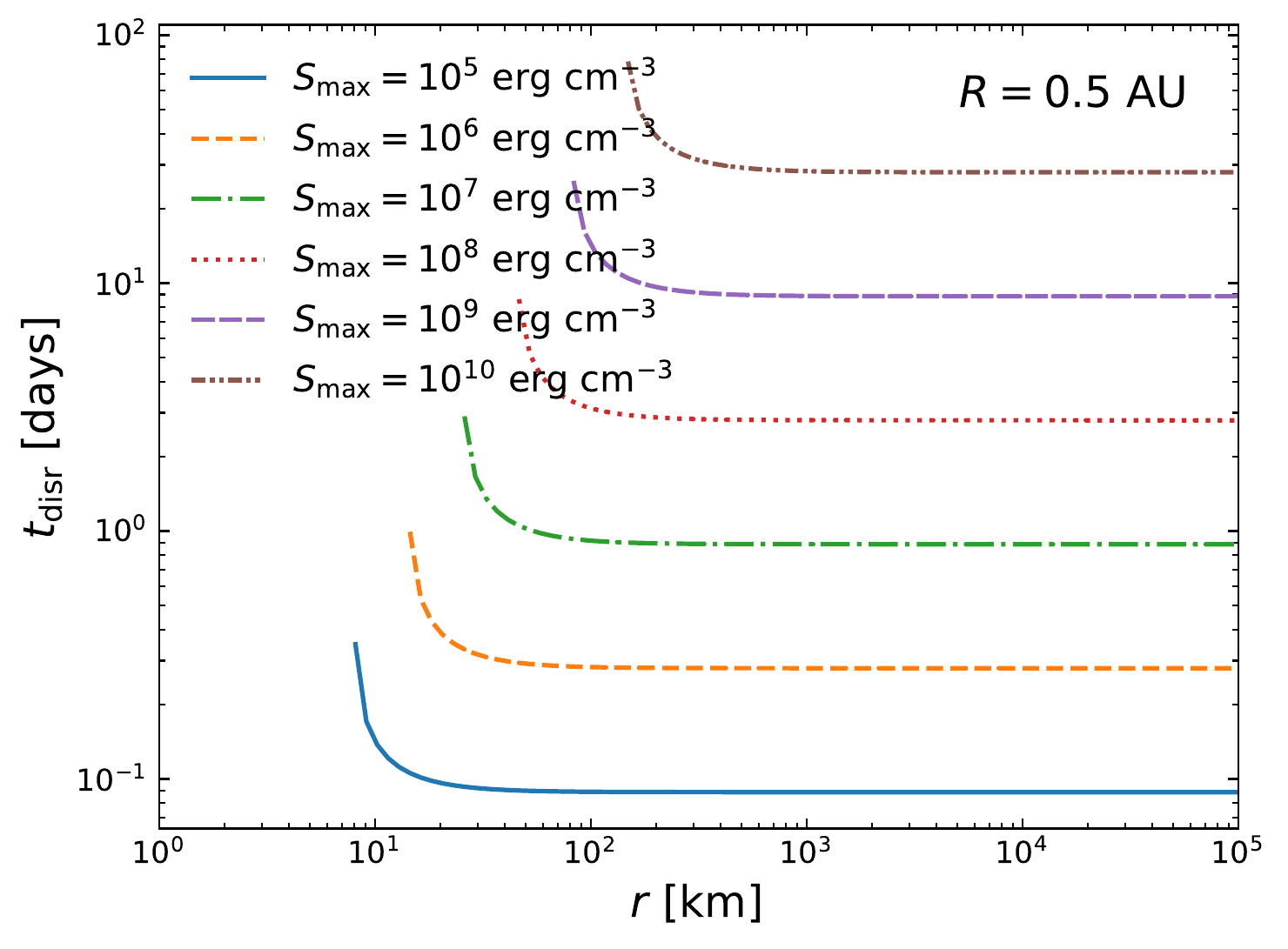}
\includegraphics[width=0.5\textwidth]{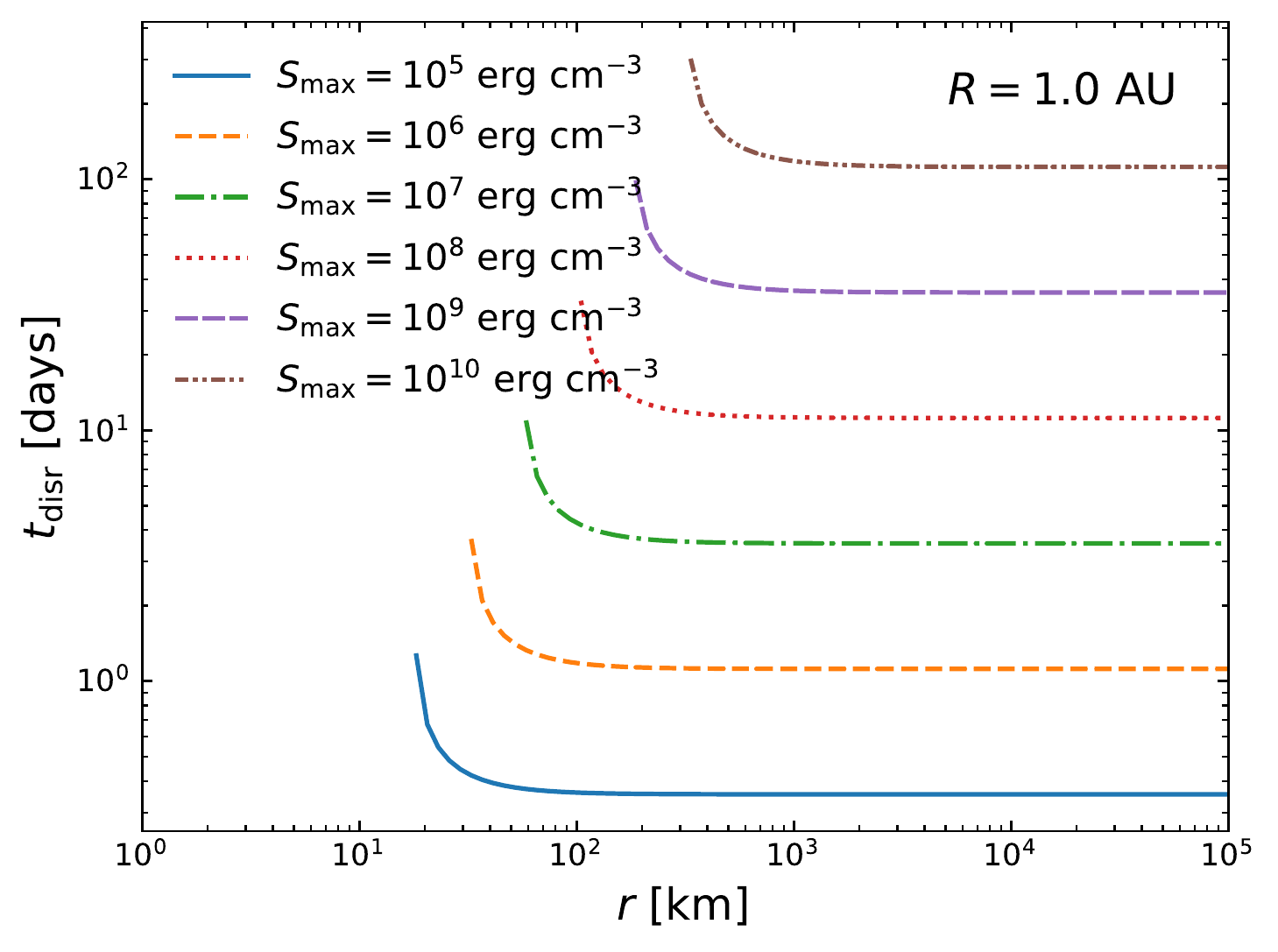}
\includegraphics[width=0.5\textwidth]{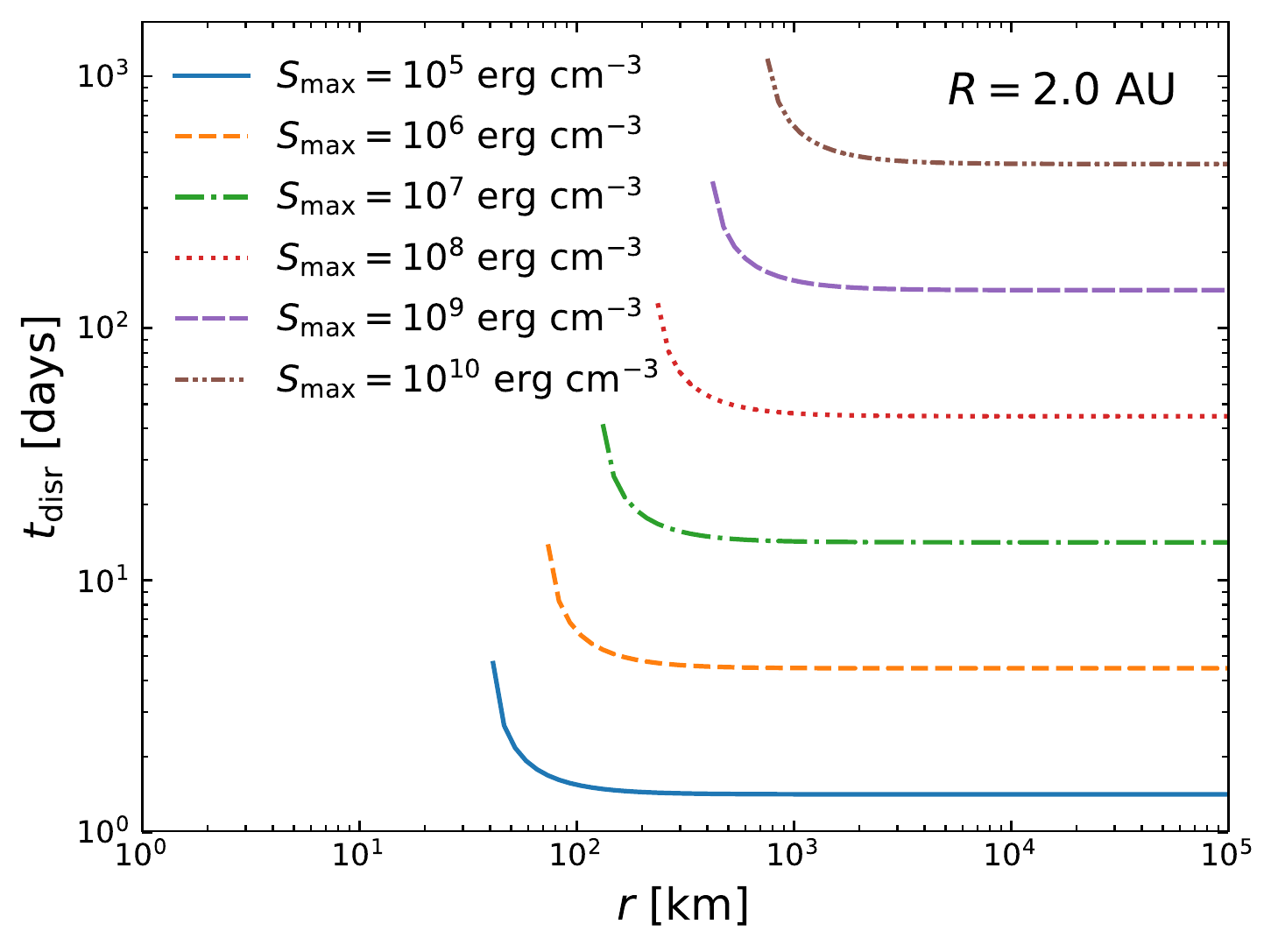}
\caption{Same as Figure \ref{fig:adisr_Qconst}, but for the disruption time of $a=1\mum$ grains. The disruption time decreases with decreasing tensile strength $S_{\max}$ and heliocentric distance $R$.}
\label{fig:tdisr_Qconst}
\end{figure*}

Figure \ref{fig:tdisr_Qconst} shows the disruption time as a function of the cometocentric distance $r$ where the disruption process starts, for different tensile strengths, $S_{\max}$, and different heliocentric distances, $R$, assuming a constant $Q_{\rm gas}$. For a given $R$, the disruption time first rapidly decreases with $r$ and quickly becomes saturated at large $r$ where the IR damping becomes dominant over the gas damping. The disruption time increases with increasing $S_{\max}$. For instance, the disruption time is rather short, of $t_{\rm disr}\lesssim 10$ days at $R=1\AU$, and decreases to $t<1$ days at $R=0.1$\AU, assuming $S_{\max}\lesssim 10^{8}\erg\cm^{-3}$. Here we neglect the motions of dust particles throughout the coma for simplicity. In reality, as the particle moves outward to a larger cometocentric distance and experiences greater RATs which induced faster rotation, $t_{\rm disr}$ could be much smaller accordingly.

\begin{figure*}
\includegraphics[width=0.5\textwidth]{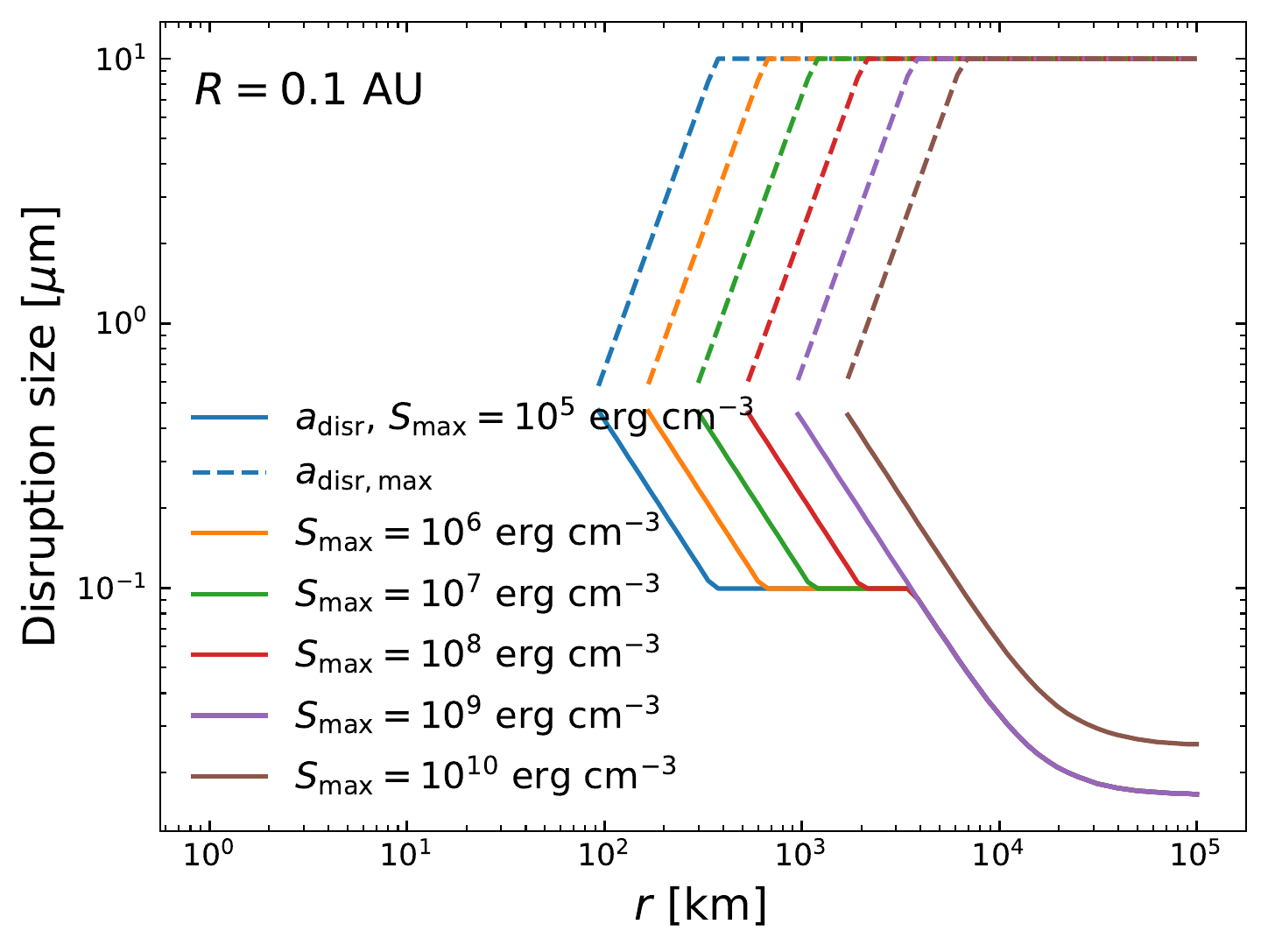}
\includegraphics[width=0.5\textwidth]{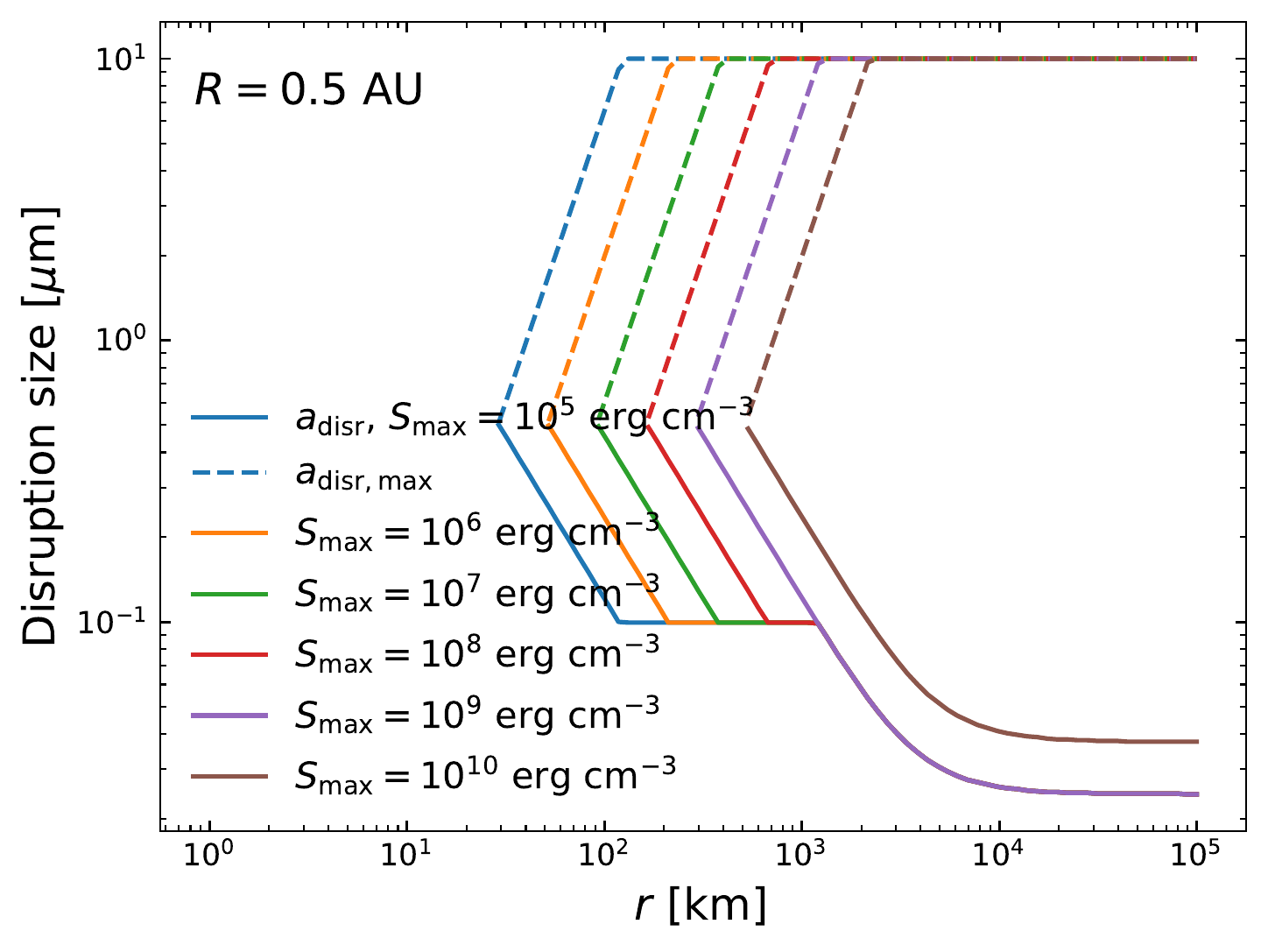}
\includegraphics[width=0.5\textwidth]{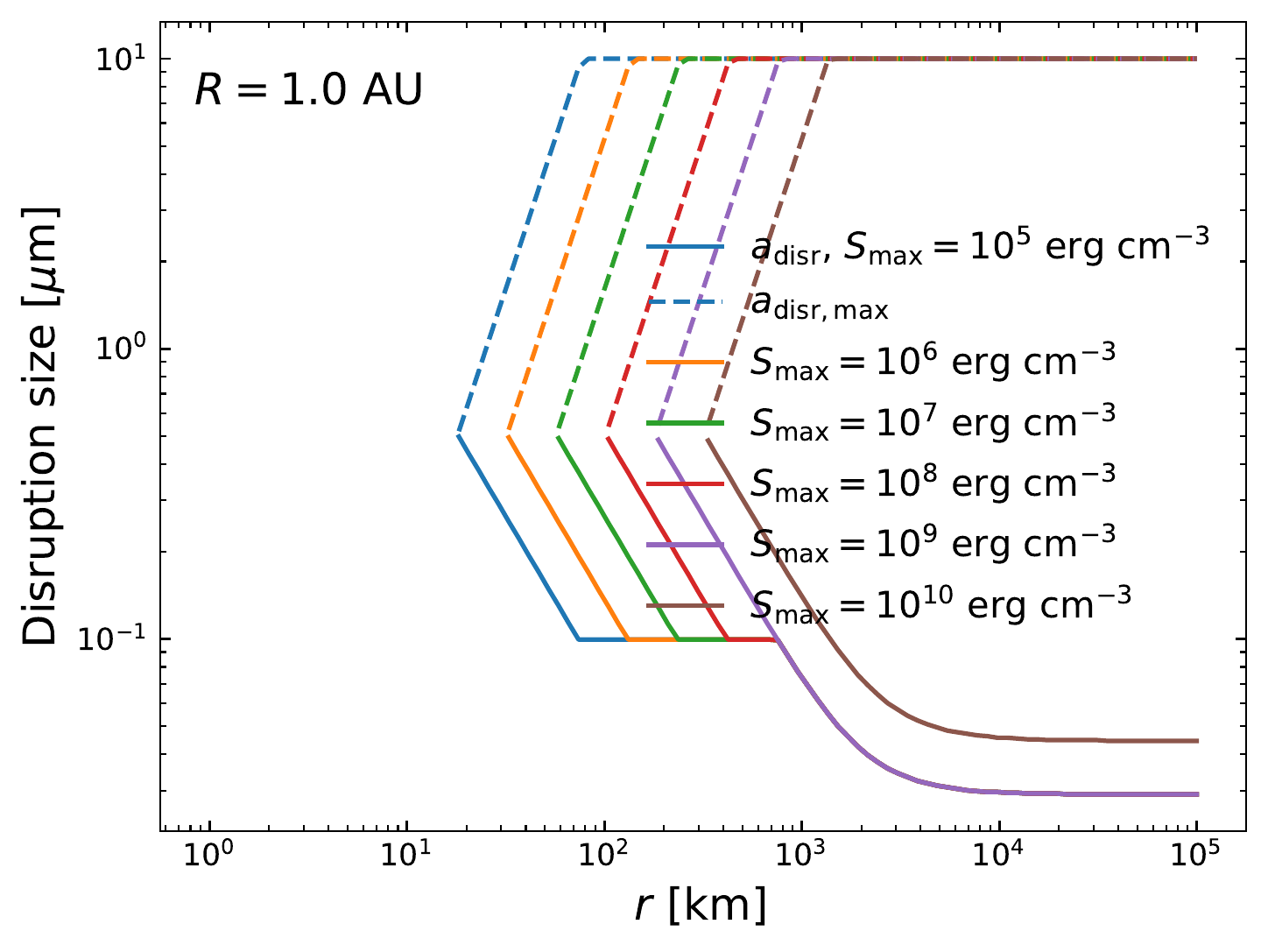}
\includegraphics[width=0.5\textwidth]{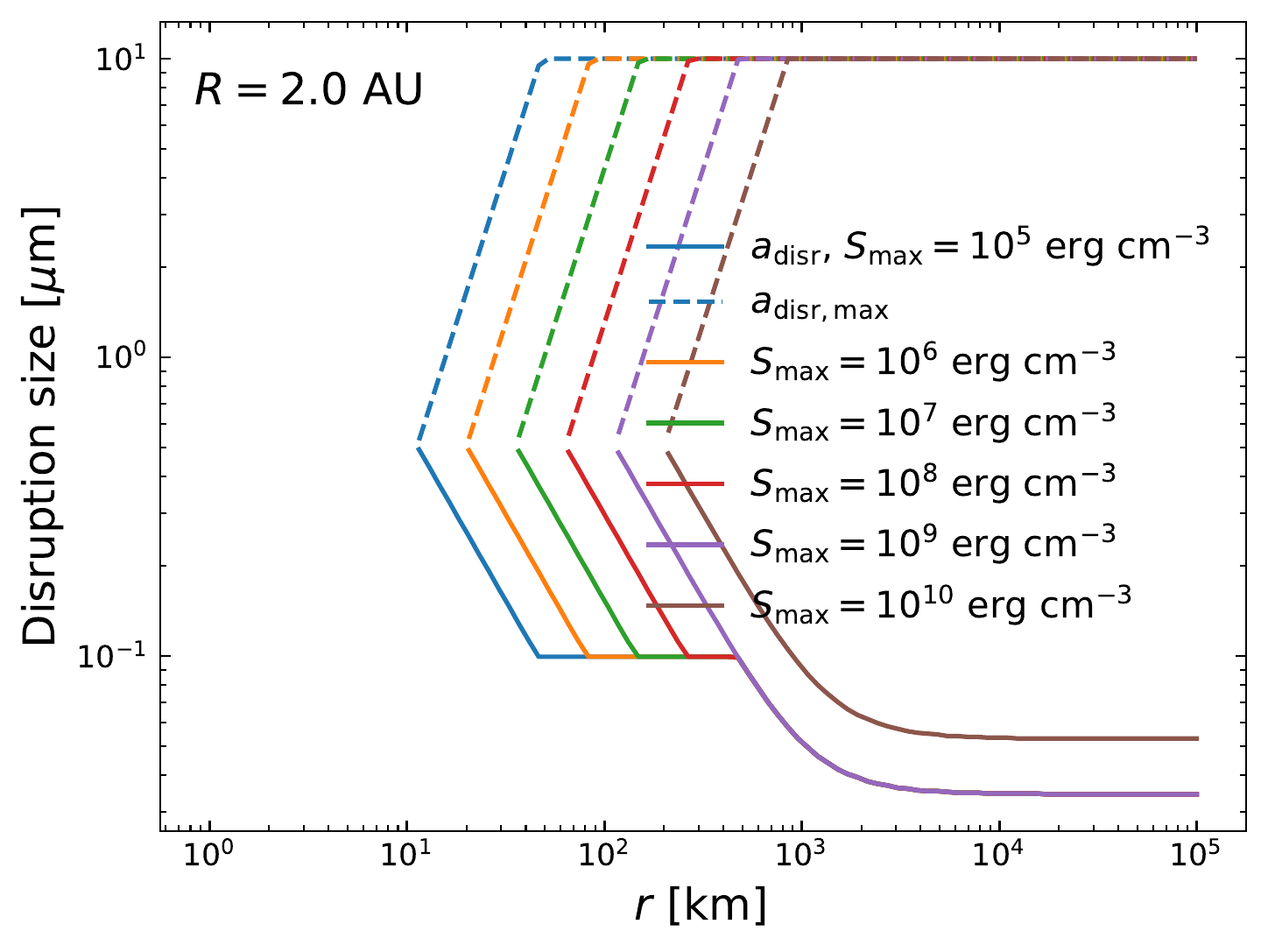}

\caption{Grain disruption size vs. cometocentric distance, $r$, for different tensile strengths, assuming varying $Q_{\rm gas}$. Disruption zone is more extended for larger $R$ due to lower $Q_{\gas}$.}
\label{fig:adisr_Qvary}
\end{figure*}

\begin{figure*}
\includegraphics[width=0.5\textwidth]{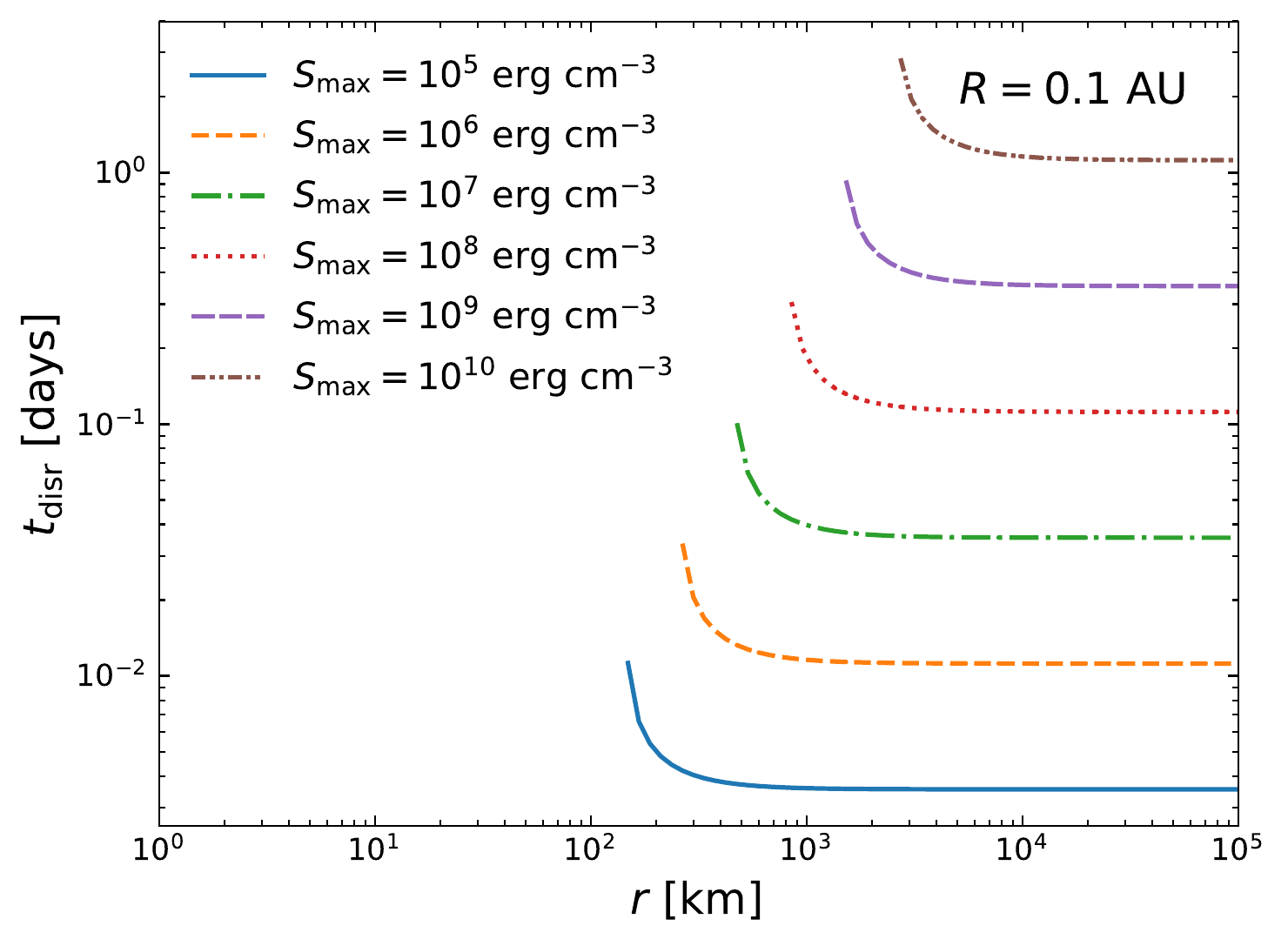}
\includegraphics[width=0.5\textwidth]{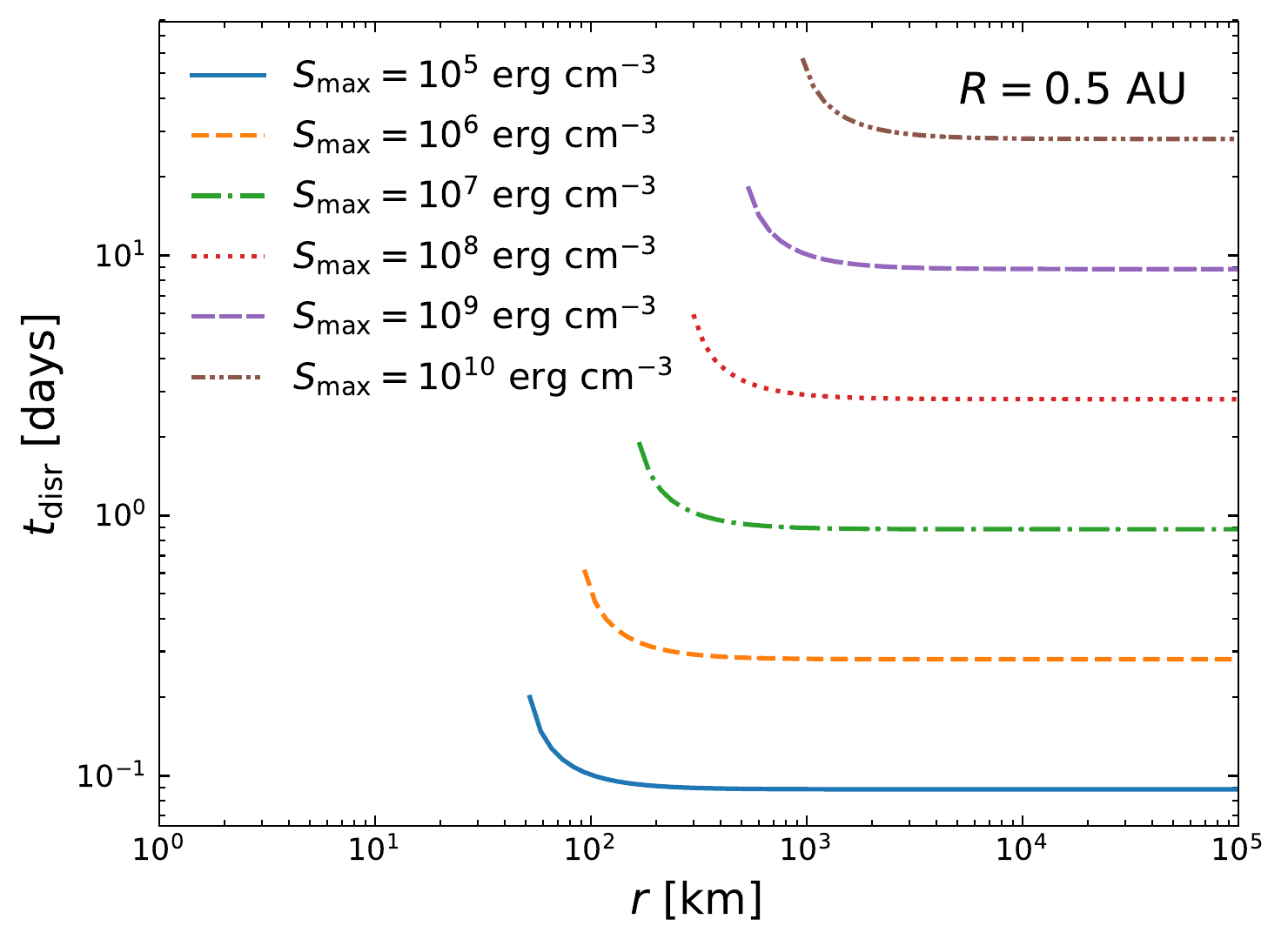}
\includegraphics[width=0.5\textwidth]{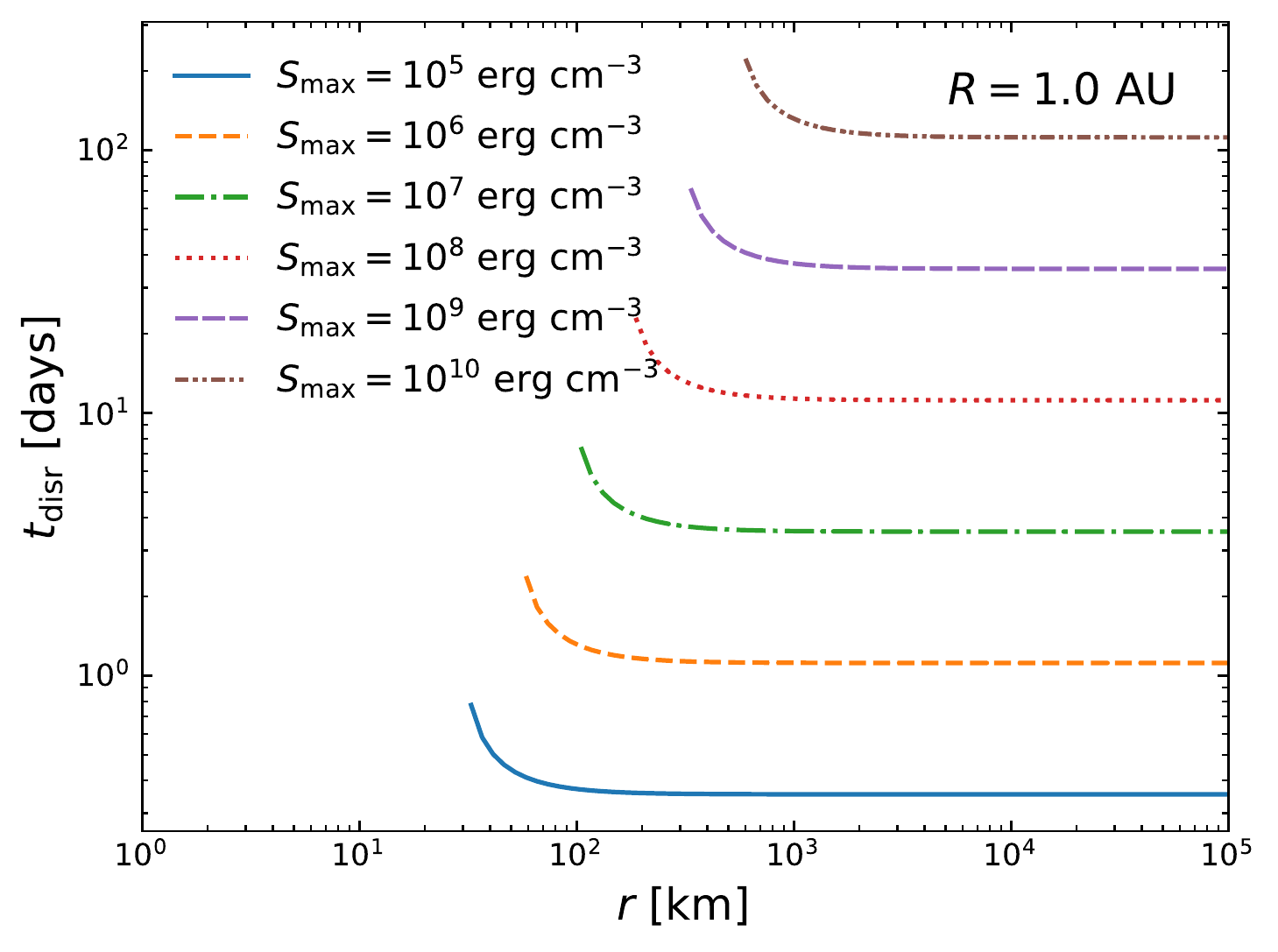}
\includegraphics[width=0.5\textwidth]{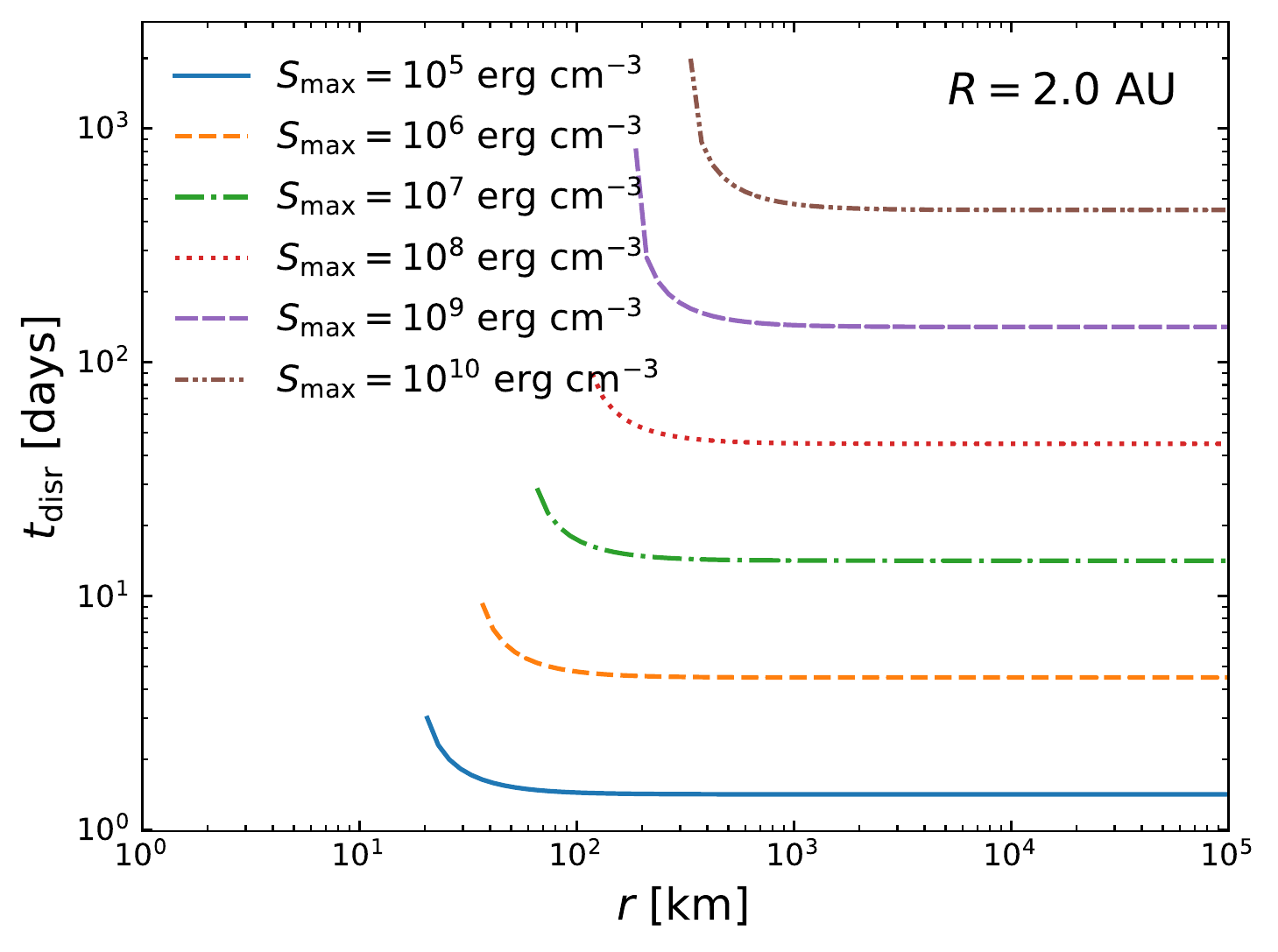}
\caption{Same as Figure \ref{fig:tdisr_Qconst} but for varying $Q_{\rm gas}$. Disruption takes place within the same timescale but at larger cometocentric distances $r$ for small heliodistances ($R = 0.1, 0.5 \AU$) and at smaller $r$ for large $R$ ($1, 2\AU$).}
\label{fig:tdisr_Qvary}
\end{figure*}

\begin{figure*}
\includegraphics[width=0.5\textwidth]{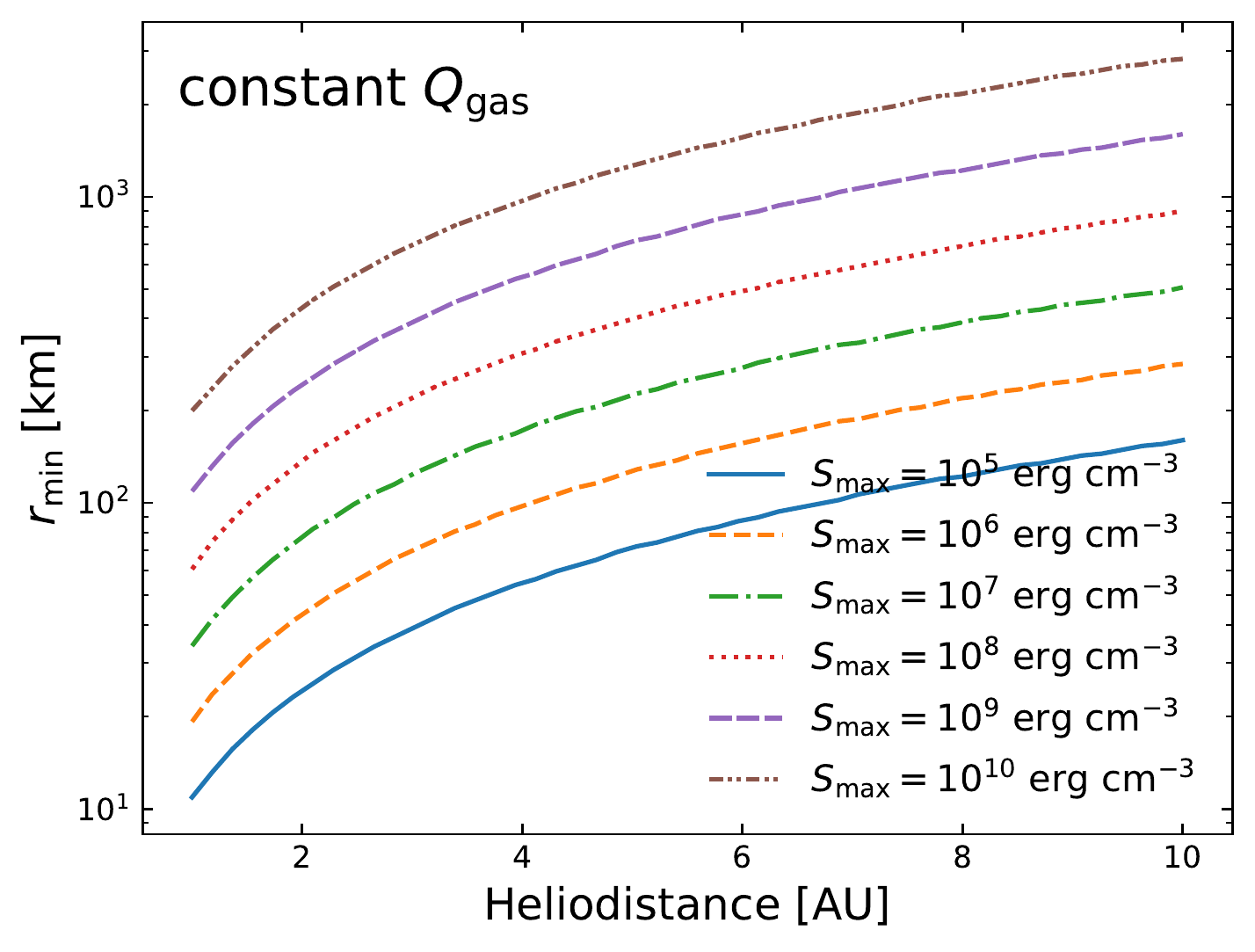}
\includegraphics[width=0.5\textwidth]{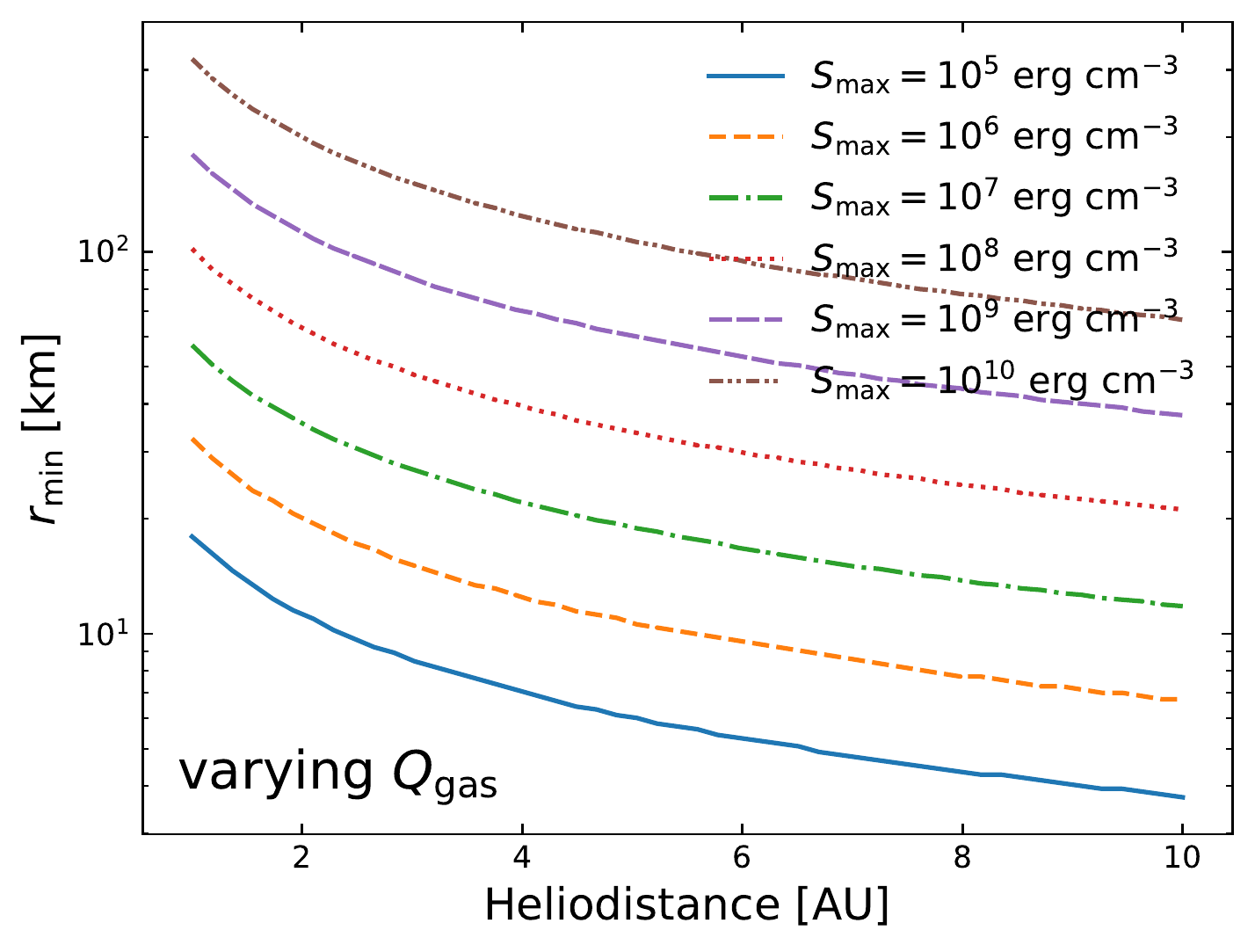}
\caption{Minimum cometocentric distance where rotational disruption occurs, $r_{\rm min}$, as a function of the heliodistance $R$ for two cases of constant $Q_{\rm gas}$ (left panel) and varying $Q_{\rm gas}$ (right panel). $r_{\min}$ increases with increasing $R$ and $S_{\max}$.}
\label{fig:rmin_disr}
\end{figure*}

To account for the dependence of $Q_{\rm gas}$ on the heliocentric distance, we calculate the disruption size for $Q_{\rm gas}$ described by Equation (\ref{eq:Qgas}). The results are shown in Figures \ref{fig:adisr_Qvary} and \ref{fig:tdisr_Qvary}. The variations of $a_{\rm disr}$ and $a_{\rm disr,max}$ with cometocentric distance $r$ are similar as in the case of constant $Q_{\gas}$ (Figure \ref{fig:adisr_Qconst}). However, their variation with $R$ is radically different due to the steep decrease of $Q_{\rm gas}$ with $R$. Specifically, at small $R$, grain disruption starts to occur at a much larger cometocentric distance of $r\sim 100\km$ for $R = 0.1\AU$. On the other hand, at large distances of $R=2\AU$, grain disruption occurs at small $r$, of $r\sim 10$ km (see more details in Figure \ref{fig:rmin_disr}).

Similarly, the time it takes to disrupt grains by RATD is of the same order as in Figure \ref{fig:tdisr_Qconst} for the same heliocentric distance, but the curves are shifted to higher cometocentric distances because the high gas density prohibits the rotational disruption to occur near the nuclei.

Figure \ref{fig:rmin_disr} shows the minimum cometocentric distance where the rotational disruption occurs as a function of the heliocentric distance for the different tensile strength, assuming a constant $Q_{\rm gas}$ (left panel) and varying $Q_{\gas}$ (right panel). As expected, weak grains are disrupted even very close to the nuclei ($\sim 8\km$ at $R = 0.1\AU$ for fluffy grains with $S_{\max}=10^5\erg\cm^{-3}$), implying that they could be disrupted as soon as being lifted off by out-gassing. Grains of large tensile strengths, $S_{\rm max}\gtrsim 10^{9}\erg\cm^{-3}$, are disrupted in regions of $r>100\km$. Moreover, for the constant $Q_{\rm gas}$ (left panel), the minimum disruption cometocentric distance increases with $R$ because of the decrease of the radiation flux. For the varying $Q_{\rm gas}\propto R^{-3.7}$, $r_{\rm min}$ decreases with $R$ because the gas damping rate decreases faster than the spin-up rate by RATs (right panel). Grain disruption still occurs at large heliodistances of $R>5\AU$.

\section{Rotational desorption of cometary ice}\label{sec:ice}
We now study rotational desorption of water ice mantles from dust grains by radiative torques. Such grains are originally present in cometary nuclei and lifted off by outgassing of highly volatile ices such as CO and CO$_2$ (\citealt{2019A&A...630A..33H}; \citealt{2016Icar..277...78F}).

\subsection{Rotational desorption of icy grain mantles}
\begin{figure}
\includegraphics[width=0.5\textwidth]{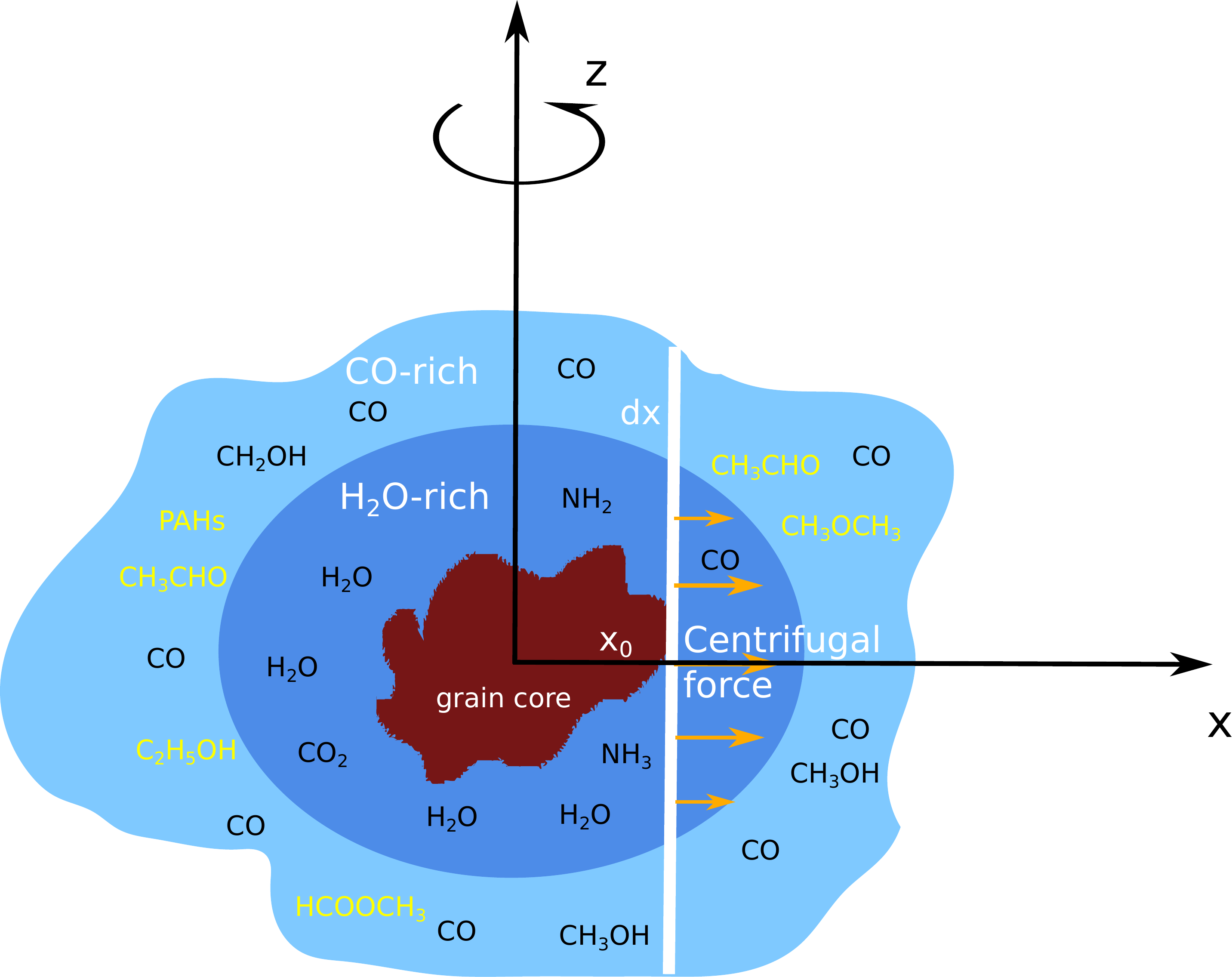}
\caption{Schematic illustration of a rapidly spinning core-mantle grain of irregular shape, comprising an icy water-rich (blue) and CO-rich (orange) mantle layers. The core is assumed to be compact silicate material, and complex organic molecules are formed in the ice mantle of the core. Centrifugal force field on a slab $dx$ is illustrated, which acts to pull off the ice mantle from the grain core at sufficiently fast rotation.}
\label{fig:icygrain}
\end{figure}

Here we consider a grain model consisting of an amorphous silicate core covered by a double-layer ice mantle (see Figure \ref{fig:icygrain}). Let $a_{c}$ be the radius of silicate core and $\Delta a_{m}$ be the average thickness of the mantle. The exact shape of icy grains is unknown, but we can assume that they have irregular shapes as required by strongly polarized H$_{2}$O and CO ice absorption features (\citealt{1996ApJ...465L..61C}; \citealt{2008ApJ...674..304W}). Thus, one can define an effective radius of the grain, $a$, which is defined as the radius of the sphere with the same volume as the grain. The effective grain size is $a \approx a_{c}+\Delta a_{m}$. The grain core is assumed to have a typical radius of $0.05 \mum$ \citep{1989IAUS..135..345G}.

The tensile strength of the bulk ice is $S_{\max}\sim 2\times 10^{7}\erg\cm^{-3}$ at low temperatures and decreases to $5\times10^{-6}\erg\cm^{-3}$ as the temperature increases to $200-300\K$ \citep{2012JGRE..117.8013L}. The adhesive strength between the ice mantle and the solid surface has a wide range, depending on the surface properties (\citealt{Itagaki:1983ud}; \citealt{Work:2018bu}).

When the rotation rate is sufficiently high such as the tensile stress exceeds the maximum limit of the ice mantle, $S_{\rm max}$, the grain is disrupted. The critical rotational velocity is determined by $S_{x}=S_{\rm max}$:
\bea
\omega_{\rm disr}&=&\frac{2}{a(1-x_{0}^{2}/a^{2})^{1/2}}\left(\frac{S_{\max}}{\rho_{\rm ice}} \right)^{1/2}\nonumber\\
&\simeq& \frac{6.3\times 10^{8}}{a_{-5}(1-x_{0}^{2}/a^{2})^{1/2}}\hat{\rho}_{\rm ice}^{-1/2}S_{\max,7}^{1/2}~\rad\s^{-1},\label{eq:omega_disr}
\ena
where $x_{0}$ is the distance from the core-mantle interface to the spinning axis (see Figure \ref{fig:icygrain}), $\rho_{\rm ice}=1\g\cm^{-3}$ is the mass density of ice, and $\hat{\rho}_{\rm ice}=\rho_{\rm ice}/(1\g\cm^{-3})$.

Above, we assume that the grain is spinning along the principal axis of maximum inertia moment. This assumption is valid because internal relaxation within the rapidly spinning grain due to Barnett effect rapidly brings the grain axis to be aligned with its angular momentum (\citealt{1979ApJ...231..404P}; \citealt{1999MNRAS.305..615R}).

The grain disruption size of ice mantles is given by
\bea
a_{\rm disr}&\simeq& 0.13\gamma^{-1/1.7}\bar{\lambda}_
{0.5}(S_{\max,7}/\hat{\rho}_{\rm ice})^{1/3.4} (1+F_{\rm IR})^{1/1.7}\nonumber\\
&&\left(\frac{n_{8}T_{2}^{1/2}}{U_{7}}\right)^{1/1.7}\mum,~~~\label{eq:adisr_low}
\ena
for $a_{\rm disr}\lesssim a_{\rm trans}$ and $x_{0}\ll a$, which depends on the local gas density and temperature due to gas damping. The equation indicates that all grains in the size range $a_{\rm trans}>a>a_{\rm disr}$ would be disrupted. 

\subsection{Desorption time and lifetime of water ice grains}

In the absence of rotational damping, the characteristic timescale for rotational desorption of ice mantles can be estimated from Equation (\ref{eq:tdisr}):
\bea
t_{\rm des}=\frac{I\omega_{\rm disr}}{\Gamma_{\rm RAT}}\simeq 2.2(\gamma U_{7})^{-1}\bar{\lambda}_{0.5}^{1.7}\hat{\rho}_{\rm ice}^{1/2}S_{\max,7}^{1/2}a_{-5}^{-0.7}{~\rm days}\label{eq:tdisr_ice}
\ena
for $a_{\rm disr}<a \lesssim a_{\rm trans}$, and
\bea
t_{\rm des}\simeq& 0.14(\gamma U_{7})^{-1}\bar{\lambda}_{0.5}^{-1}\hat{\rho}_{\rm ice}^{1/2}S_{\max,7}^{1/2}a_{-5}^{2}{~\rm days}\label{eq:tdisr_ice2}
\ena
for $a_{\rm trans}<a<a_{\rm disr,max}$.

The lifetime of water ice grains is essentially the grain desorption time $t_{\rm des}$. Therefore, ice grains can survive in the coma in a time of $t_{\rm des}$.

In the case of non-rotating grains, the lifetime of ice grains is described by thermal sublimation time. For comparison, we also compute the sublimation time of the ice mantle of thickness $\Delta a_{m}$, as given by
\bea
t_{\rm sub}(T_d)=-\frac{\Delta a_{m}}{da/dt}
= \frac{\Delta a_{m}}{l\nu_{0}}\exp\left(\frac{E_{b}}{T_d}\right),\label{eq:tausub}
\ena
where $da/dt=l/\tau_{\rm evap}$ is the rate of decrease in the mantle thickness due to thermal sublimation, $l$ is the thickness of the ice monolayer, and $\tau_{\rm evap}$ is the characteristic time that molecules stay on the grain surface before evaporation:
\bea
\tau_{\rm evap}^{-1}=\nu_{0}\exp\left(\frac{-E_{b}}{T_d}\right),\label{eq:tevap}
\ena
where $\nu_{0}$ is the characteristic vibration frequency of the lattice, and $E_{b}$ is the binding energy (\citealt{1972ApJ...174..321W}).

Plugging the numerical parameters of water ice into the above equation, we obtain
\bea
t_{\rm sub}\sim 1.5\times 10^{3}\left(\frac{\Delta a_{m}}{500\AA}\right)\exp\left(\frac{E_{b}}{4800\K}\frac{100\K}{T_{d}} \right) \yr.\label{eq:tsub}
\ena

For $T_{d}<100\K$, i.e., $U\sim (T_{d}/16.4\K)^{6}\lesssim 5\times 10^{4}$, Equation (\ref{eq:tdisr_ice2}) yields the desorption time $t_{des}\sim 27$ days, much shorter than $t_{\rm sub}>1500$ yr for pure water ice given by Equation (\ref{eq:tsub}). For dust grains with ice mantles as in our model, ice may sublimate faster, but a very specific ratio of $\Delta a_{m}/a_{c}$ at a certain $T_{d}$ is required for the mantle to sublimate over the distance scale of the observable coma and thus, a steep brightness gradient in the observable coma at $R < R_{\rm sub}(\H_2\O)$ cannot be explained by thermal sublimation \citep{1981Icar...47..342H}.

\subsection{Numerical Results}
Here, we adopt a conservative value of $S_{\max}=10^{7}\erg\cm^{-3}$ for ice mantles for our numerical calculations. For the grain core, a higher value of $S_{\max}=10^{9}\erg\cm^{-3}$ is adopted.

\begin{figure*}
\includegraphics[width=0.5\textwidth]{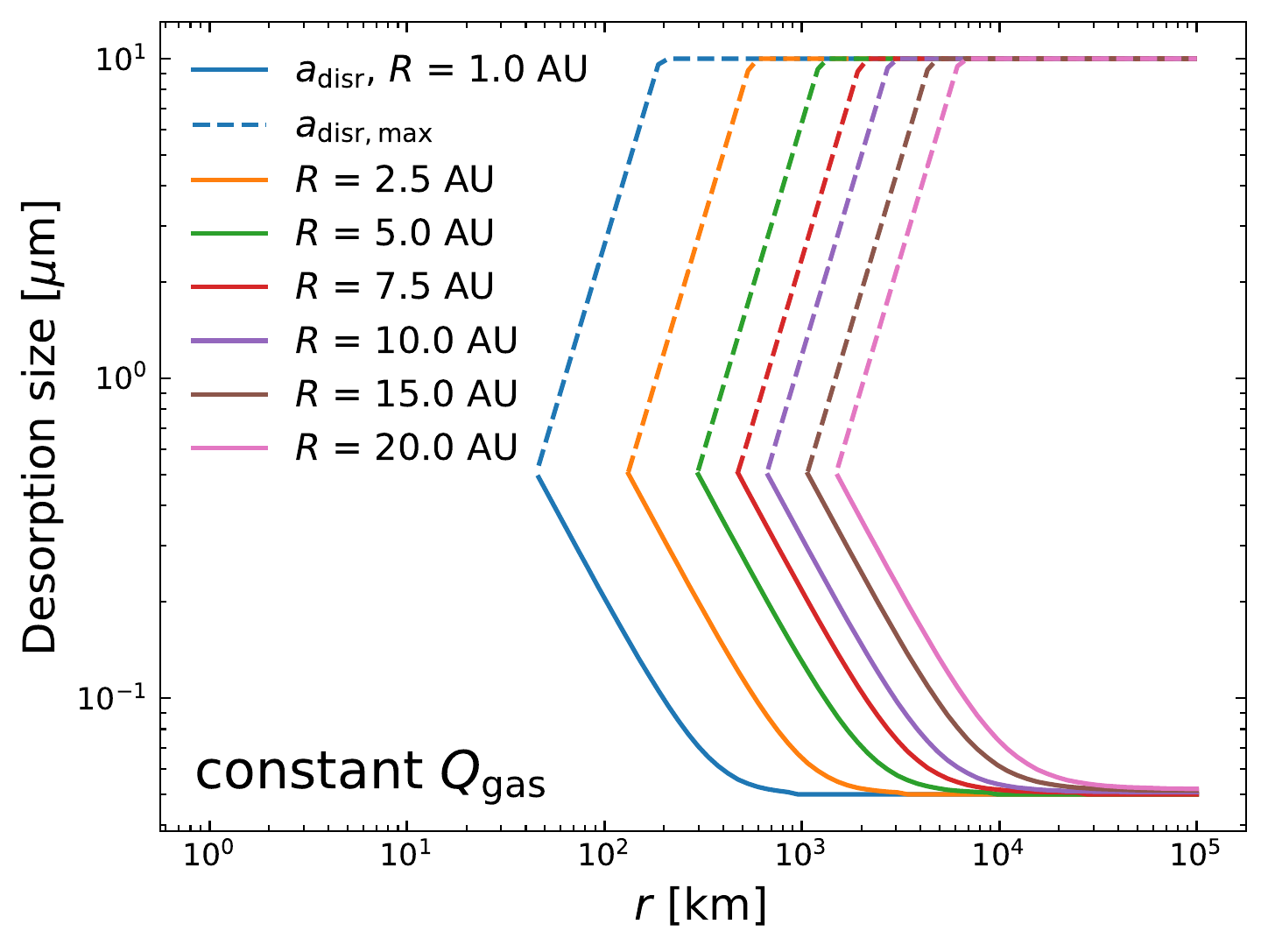}
\includegraphics[width=0.5\textwidth]{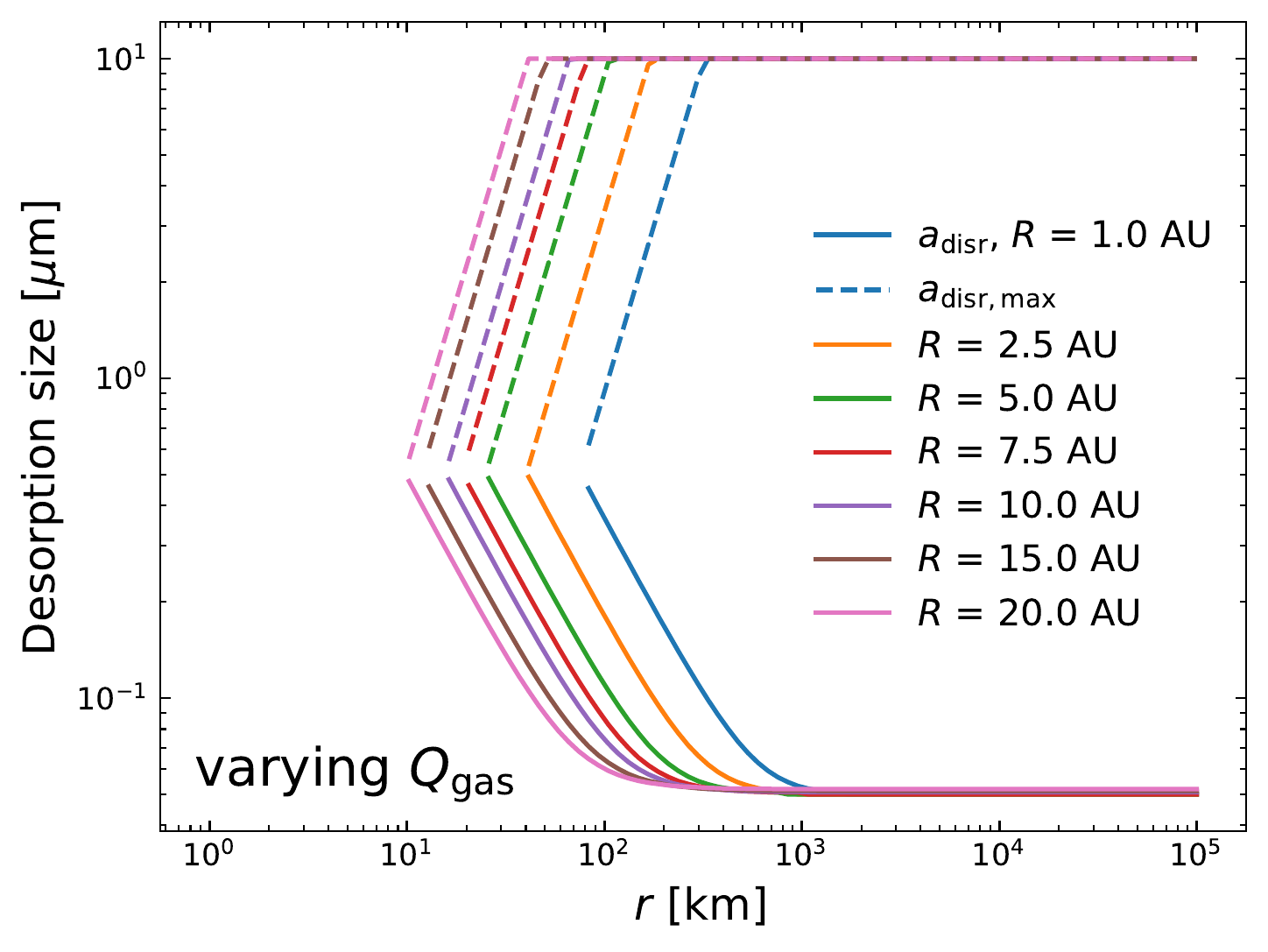}
\caption{Rotational desorption size of water ice mantles as a function of the cometocentric distance for different heliocentric distances, $R$, assuming a constant $Q_{\rm gas}$ (left panel) and varying $Q_{\gas}$ (right panel). The desorption region is more extended for smaller heliocentric distances in the cases of constant $Q_{\rm gas}$, but it is reduced when $Q_{\rm gas}$ decreases with increasing $R$.}
\label{fig:ades_ice}
\end{figure*}

Figure \ref{fig:ades_ice} shows the desorption size of mantles from the grain core, assuming $a_{c}=0.05\mum$. Rotational desorption occurs at large distance of $R\sim 5$ AU. The inner cometocentric distance of rotational desorption decreases with $Q_{\rm gas}$ and could reach $r_{\rm des,min}\sim 1\km$ for $Q_{\rm gas}\sim 10 \g\cm^{-3}$. The small $Q_{\rm gas}$ corresponds to low gas density $n_{\rm gas}(r)$, which efficiently damps grain rotation spun-up by RATs. For the constant $Q_{\rm gas}$, the desorption occur at larger cometocentric distances $r$ for larger heliocentric distance $R$. For the case of varying $Q_{\rm gas}$, rotational desorption occurs at smaller $r$ for larger $R$ because the gas production rates decreases more rapidly with $R$ than the decrease of $n_{\H}$ with $r$. 

\begin{figure}
\includegraphics[width=0.45\textwidth]{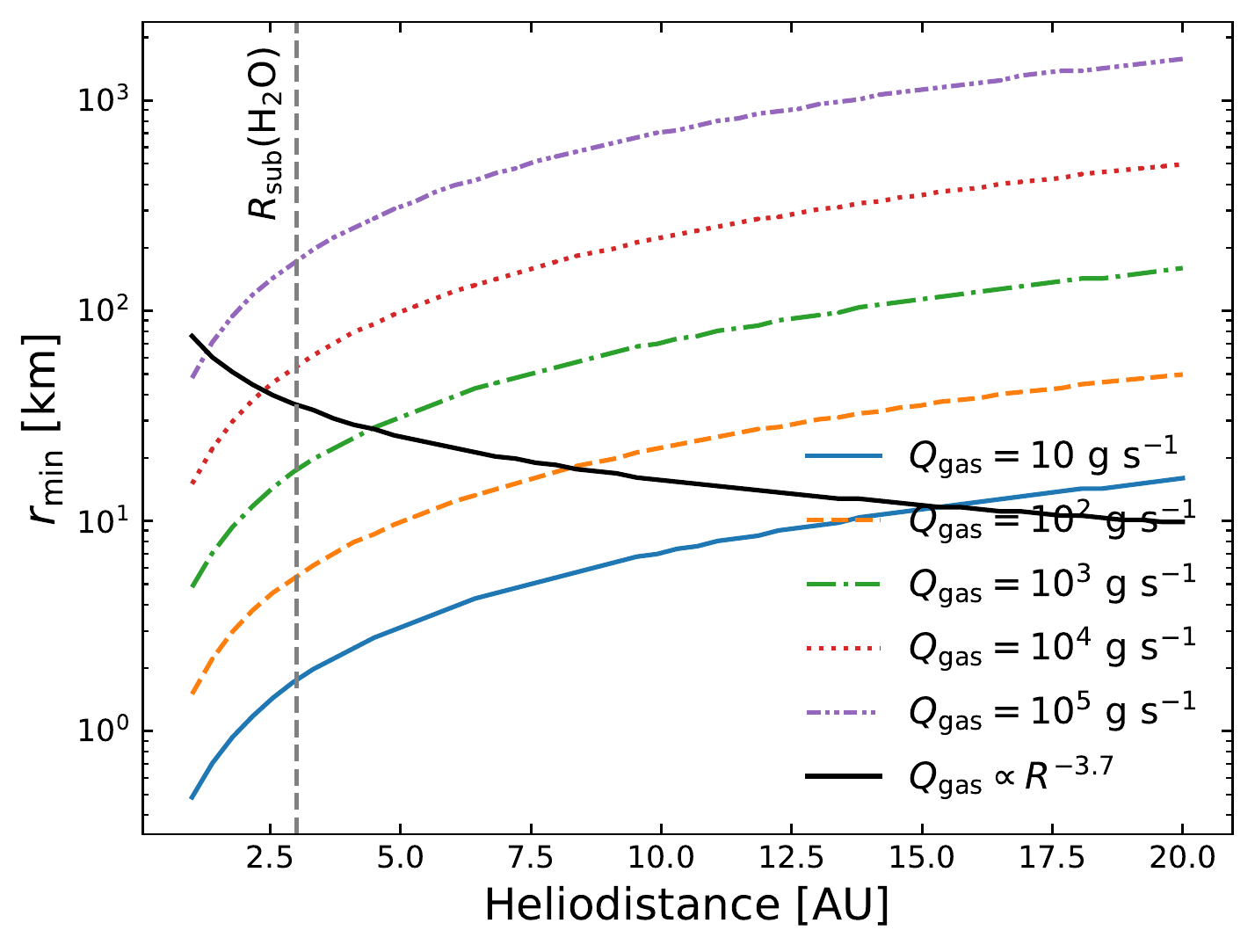}
\caption{Minimum cometocentric distance where rotational desorption occurs, $r_{\rm min}$, as a function of the heliodistance $R$ for different values of $Q_{\rm gas}$. Black solid line is the plot of $r_{\min}$ when the power-law variation of $Q_{\rm gas}$ with $R$ is considered. Rotational desorption could occur well beyond the sublimation zone marked by the vertical dashed line.}
\label{fig:rdes_min}
\end{figure}

Figure \ref{fig:rdes_min} shows the minimum cometocentric distance for which rotational desorption begins to occur $r_{\min}$ as a function of the heliodistance for different values of $Q_{\rm gas}$. At a large distance from the Sun, rotational desorption of ice mantles takes place at a greater cometocentric distance due to the lower radiation flux. For a larger $Q_{\rm gas}$, the gas density within the coma is higher, thus it is harder to desorb ice mantles, increasing the value of $r_{\min}$.

\begin{figure*}
\includegraphics[width=0.5\textwidth]{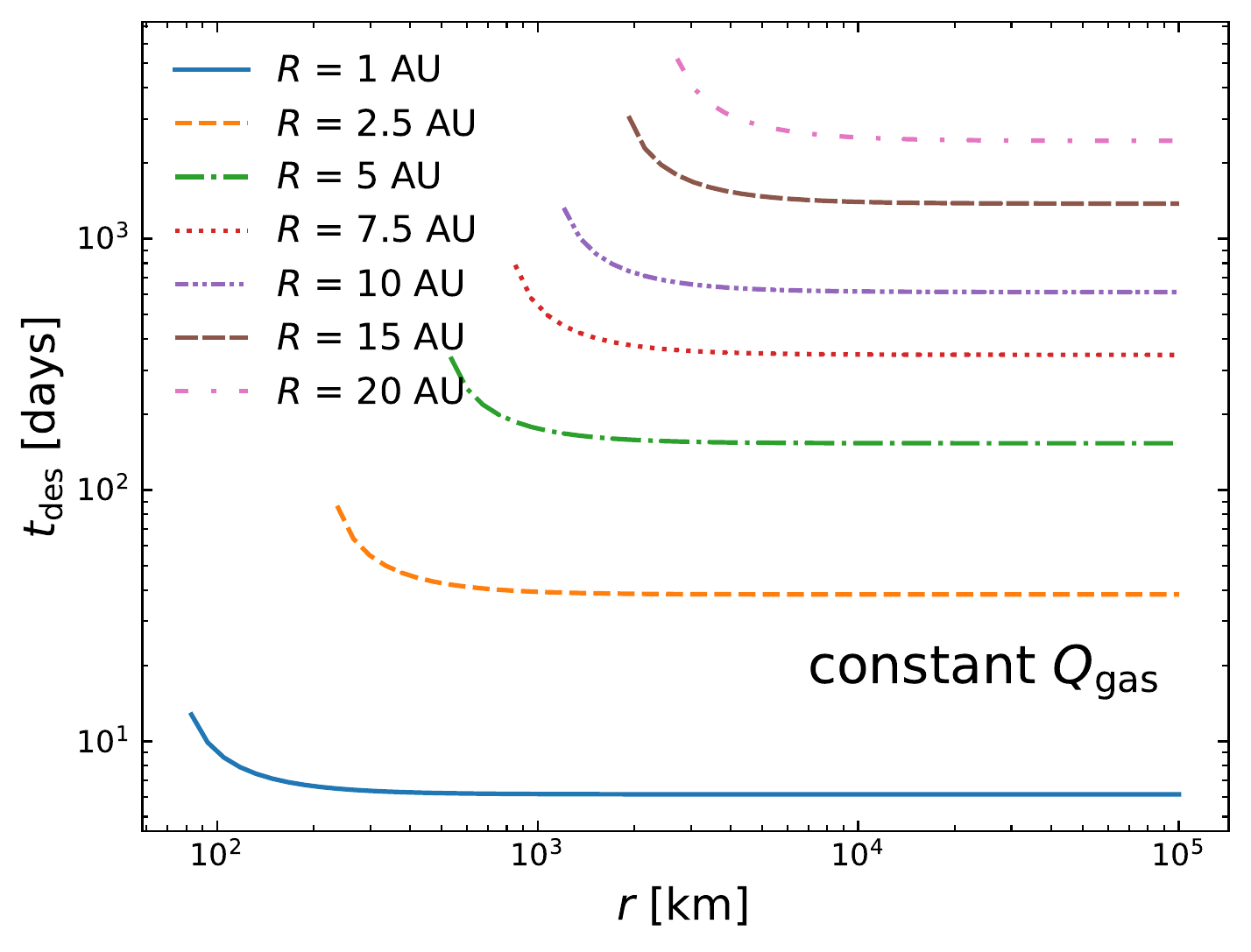}
\includegraphics[width=0.5\textwidth]{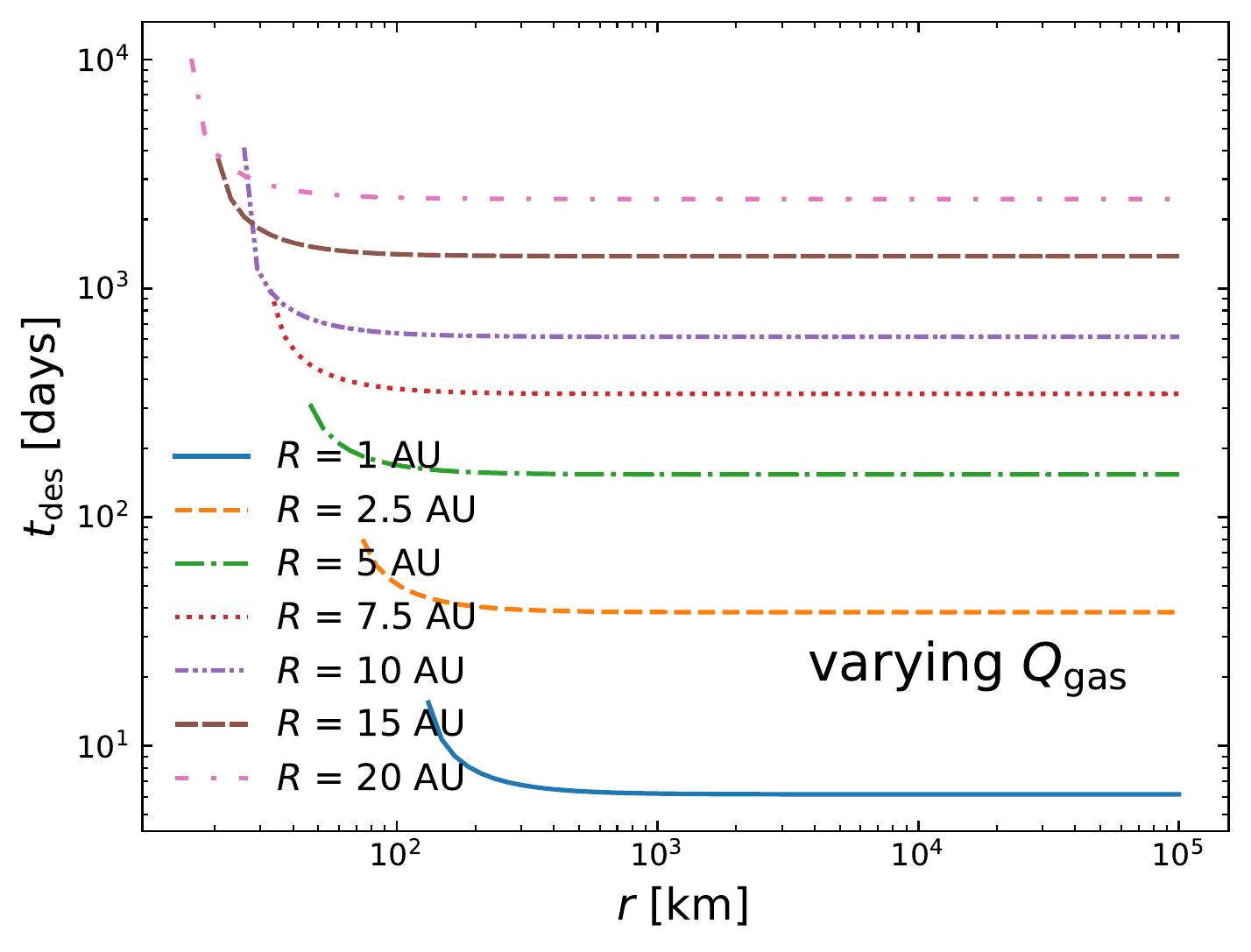}
\caption{Rotational desorption time of water ice mantles as a function of the cometocentric distance, $r$, for different heliocentric distances $R$, assuming a constant $Q_{\rm gas}$ (left panel) and varying $Q_{\rm gas}$ (right panel). Near the Sun, ice grains are desorbed rapidly, while at large $R$, ice mantles are removed in more than 100 days. Rotational desorption takes place within the same timescale in the two cases, but at different cometocentric distances due to the different variation of the gas density with $R$.}
\label{fig:tdes_ice}
\end{figure*}

Figure \ref{fig:tdes_ice} shows the desorption time of ice mantles from the grain core for the different heliocentric distances. The desorption time increases with increasing $R$ due to the decrease of the radiation flux. The desorption time rapidly decreases with $r$ and achieves a saturated value when the IR damping becomes dominant, similar to the case of grain disruption (see Figure \ref{fig:tdisr_Qconst}). 

Comet nuclei may also contain large dense or porous aggregates of $a>10\mum$, which are assembles of original icy grains \citep{Protopapa:2014}. The disruption of these aggregates into individual icy grains are similar to that of the composite grains in Section \ref{sec:dust}, followed by the desorption of the ice mantles into small fragments.

\subsection{Production rate of water vapor via rotational desorption}
The production rate of dust and water ice grains is at most equal to $Q_{\rm gas}$ because outgassing can at most lift the equal amount of dust mass. Rotational desorption of ice mantles converts water ice into water vapor. The water vapor production rate, $Q_{\H_{2}\O}$, is then determined by the mass of water ice produced by desorption.

The water ice production rate due to outgassing is given by
\bea
Q_{\rm ice}= \frac{4\pi r^{2} dr M_{\rm ice}(r)}{dt} = 4\pi r^{2}vM_{\rm ice}(r),
\ena
where the mass of ice grains 
\bea
M_{\rm ice}(r)&=&\int_{a_{\rm min}}^{a_{\rm max}} \frac{4\pi}{3} [a_{c}^{3}\rho_{c} + (a^3 - a_{c}^{3})\rho_{\rm ice}]\frac{dn_{\rm ice}}{da}da\nonumber\\
&\approx&\int_{a_{\rm min}}^{a_{\rm max}} \left(\frac{4\pi a^{3}\rho_{\rm ice}}{3}\right)\frac{dn_{\rm ice}}{da}da,
\ena
where the size distribution of individual core-ice mantle grains $dn_{\rm ice}/da= Ca^{-3.5}$ with $C$ being the normalization constant, $a_{\rm min}=0.05\mum$ and $a_{\max} = 10\mum$ as previously assumed for the calculations of rotational disruption of composite grains. Here the contribution of the grain core to the total mass is negligible because $a_{c}\ll a_{\max}$.


Then, we calculate the mass of water ice desorbed or water vapor produced as follows:
\bea
\Delta M_{\H_{2}\O}(r) = f_{\rm high-J} \int_{a_{\rm disr}}^{a_{\rm disr,max}} \rho_{\rm ice} V_{\rm ice}(a) \left(\frac{dn_{\rm ice}}{da}\right)da,~~~~
\ena
where $f_{\rm high-J}$ is the fraction of grains in the size range $[a_{\min},a_{\max}]$ that are aligned with high-J attractors by RATs (\citealt{2014MNRAS.438..680H}) and $V_{\rm ice}(a) = 4\pi (a^{3}-a_c^3)/3$ is the volume of the ice
mantle of grain with size $a$ and core radius $a_{c}$. The exact value of $f_{\rm high-J}$ depends on the grain shape and size, and one expects $0<f_{\rm high-J}<1$ for ordinary paramagnetic grains (\citealt{Herranen:2020im}) and $f_{\rm high-J}=1$ for grains with iron inclusions (\citealt{2016ApJ...831..159H}). 

The mass fraction of ice grains desorbed by rotational desorption at cometocentric distance $r$ is given by
\bea
f_{\rm des,coma}(r) &=& \frac{\Delta M_{\H_{2}\O}(r)}{M_{\rm ice}(r)}.\label{eq:fdes_r}
\ena

The total fraction of ice mass removed from the entire coma as a function of the heliocentric distance is obtained by integrating over the cometary coma $r$:
\bea
f_{\rm des,helio}(R) = \frac{\int_{r_{\min}}^{r_{\max}} \Delta M_{\H_{2}\O}(r) 4\pi r^2 dr}{\int_{r_{\min}}^{r_{\max}} M_{\rm ice}(r) 4\pi r^2 dr}.
\ena

\begin{figure*}
\includegraphics[width=0.5\textwidth]{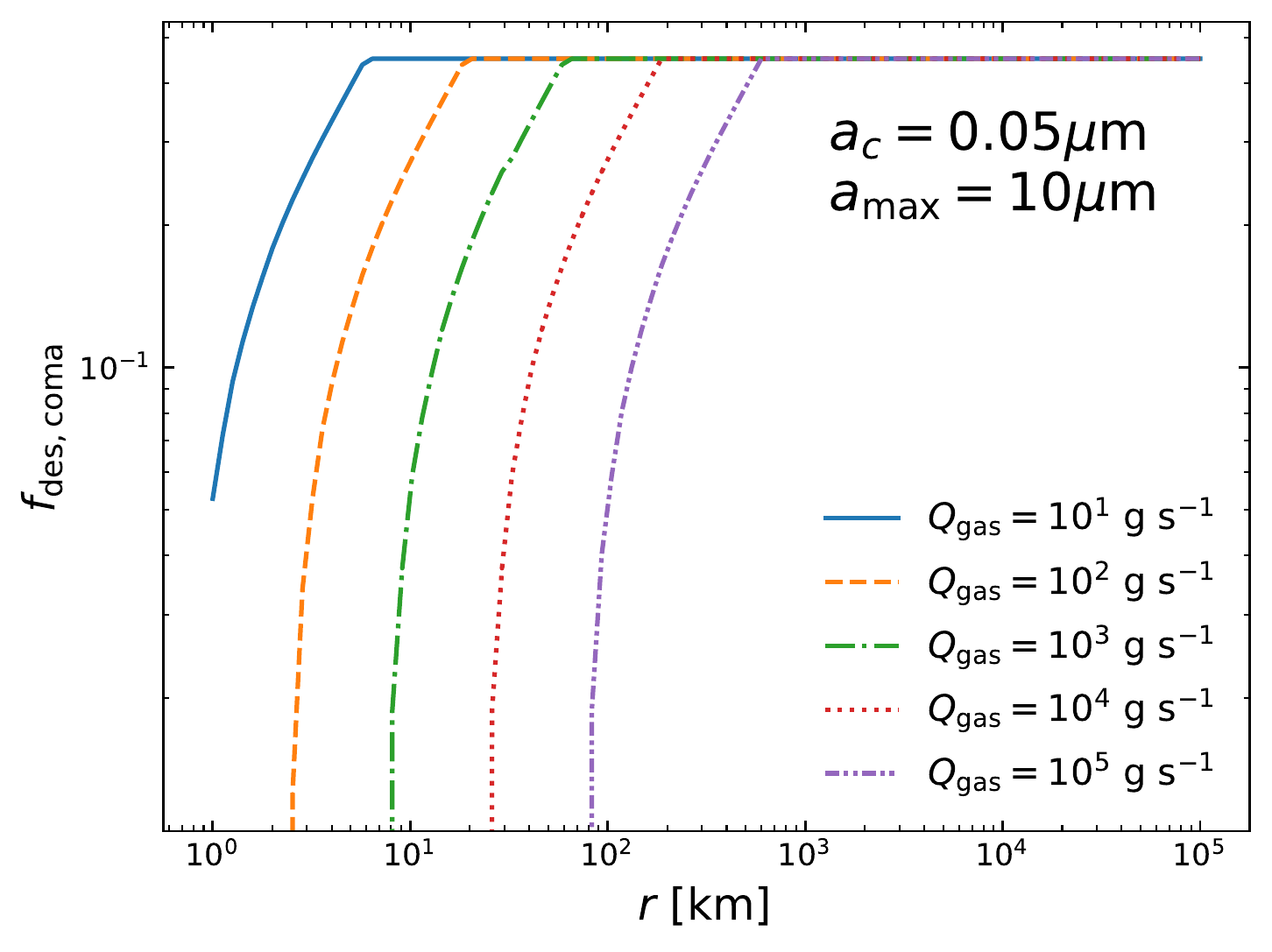}
\includegraphics[width=0.5\textwidth]{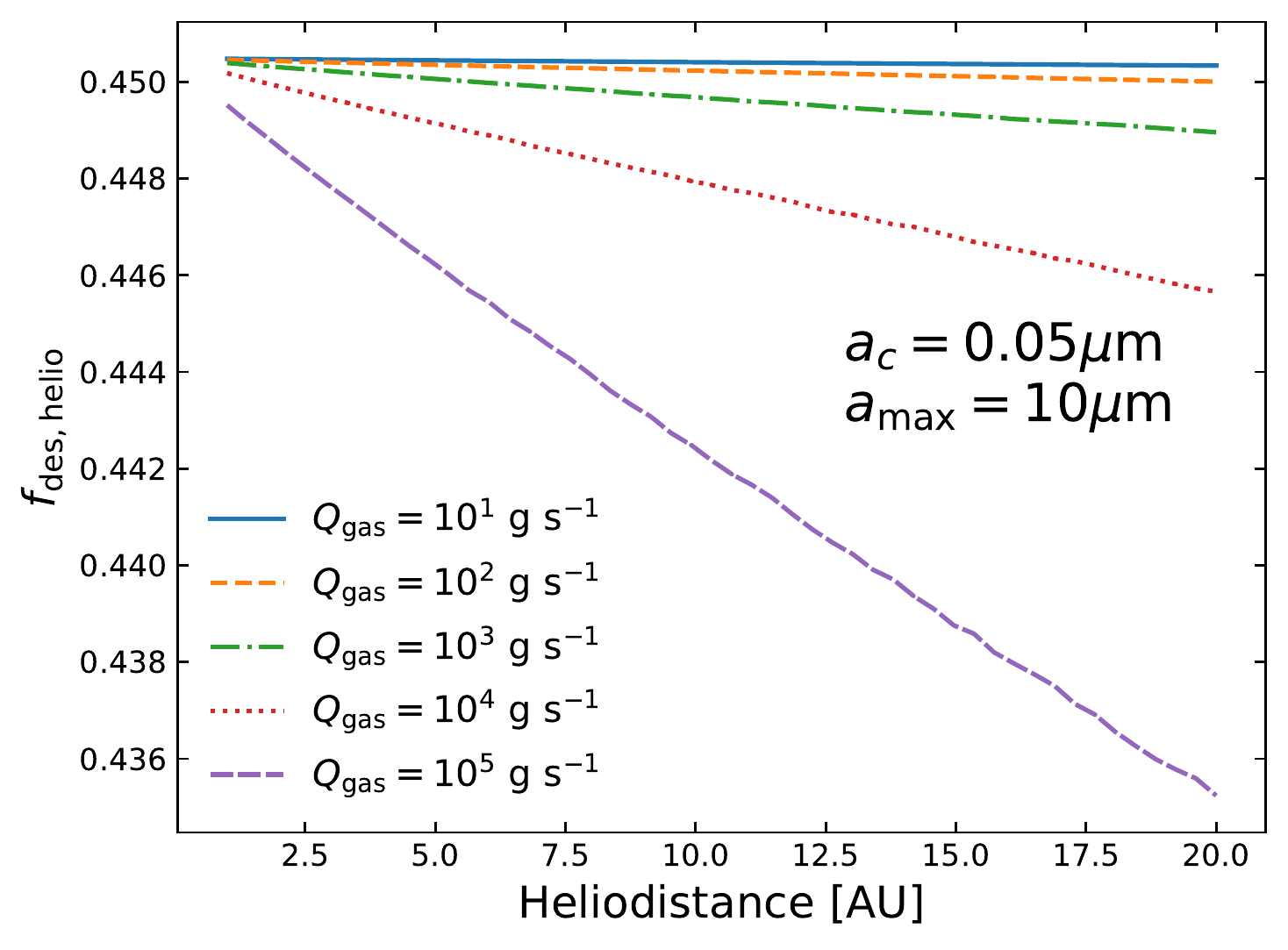}
\caption{Left panel: Fraction of ice mass removed by rotational desorption, $f_{\rm des,coma}$, within the coma as a function of the cometocentric distance $r$. $f_{\rm des,coma}$ increases rapidly with $r$ and reaches maximum of $\sim f_{\rm high-J}$. Right panel: Fraction of ice mass removed by rotational desorption $f_{\rm des,helio}$ as a function of the heliocentric distance $R$. $f_{\rm des, helio}$ slightly decreases with increasing $R$ and $Q_{\rm gas}$. Here $a_{\rm max}=10\mum$ is considered.}
\label{fig:fdes1}
\end{figure*}

Figures \ref{fig:fdes1} shows $f_{\rm des,coma}$ with the distance from the nucleus $r$ (left panel) and $f_{\rm des,helio}$ (right panel), assuming a conservative value of $f_{\rm high-J} = 0.5$ and the maximum grain size $a_{\max} = 10\mum$. The core radius is again fixed to $a_c = 0.05\mum$. The amount of ice desorbed $\Delta M_{\H_{2}\O}$ rapidly reaches maximum $\sim M_{\rm ice}/2$ at large cometocentric distance due to the disruption of the very thick mantles of large grains ($\sim 45\%$ of the ice mass). 

\section{Discussion}\label{sec:discuss}

\subsection{Implications for varying dust properties in cometary comae}
Dust grains are lifted off the comet nucleus by the outgassing of highly volatile molecules. Such grains are presumably thought to have a fluffy structure,
made of individual monomers that are loosely bound together via Van der Waals force. Theoretical calculations for the tensile strength of fluffy grains gives $S_{\rm max}\sim 10^{4}(a_{p}/0.05\mum)^{-2}$ (\citealt{2019ApJ...876...13H}) where $a_{p}$ is the monomer's radius. Experimental measurements in \cite{Gundlach:2018cu} yields $S_{\rm exp}\simeq 2.73\times 10^{5}(0.1\mum/a_{p})$. Thus, one expects a low tensile strength, i.e., $S_{\max}>10^{5}\erg\cm^{-3}$ for larger grains of fluffy structures.

In this paper, using the RATD mechanism, we study the evolution of composite grains assuming different tensile strengths of $S_{\max}=10^{5}-10^{10}\erg\cm^{-3}$ for large grains, where the maximum considered values of $S_{\rm max}\gtrsim 10^{9}\erg\cm^{-3}$ are expected for small grains (i.e., $a<0.1\mum$) of compact structures. We find that RATD is efficient in disrupting large composite grains ($a > 10\mum$) into smaller ones. Very large grains (VLGs) of fluffy structures of $S_{\max}\lesssim 10^{5}\erg\cm^{-3}$ are disrupted as soon as being released from the nucleus by outgassing. Subsequently, smaller grains of higher $S_{\rm max}$ are disrupted at larger $r$ (see Figures \ref{fig:adisr_Qconst} and \ref{fig:tdisr_Qconst}). Note that for ordinary paramagnetic or diamagnetic grains, not all large grains are disrupted, but only a fraction of such large grains ($a>a_{\rm disr}$) that are aligned with high-J attractors, $f_{\rm high-J}$, could be disrupted (see \citealt{Hoang:2020} for a review). Therefore, the RATD mechanism implies the evolution of dust properties (e.g., grain size distribution and structure), with the decrease (increase) in the abundance of large (small) grains, that depends on the cometocentric and heliocentric distances. Moreover, the efficiency of RATD increases with decreasing the gas production rate $Q_{\rm gas}$ because the latter determines the rotational damping of grains spun-up by RATs. 

Observations from spacecraft, as well as polarimetric observations indeed, reveal the variation of dust properties within cometary comae. In-situ measurements by spacecraft report the fragmentation of dust grains on encounter of Comet Haley after released from nucleus (\citealt{1986Natur.321..278S}; \citealt{1986Natur.321..336K}; \citealt{2004Sci...304.1776T}). In-situ spacecraft measurements at a distance 8000 km from the nucleus shows the decrease of large grains $m>10^{-13}\g$ with distance \citep{1986Natur.321..278S}. The Stardust, COSIMA, and GIADA instruments onboard Rosetta collect dust particles lifted off  the nucleus of the Jupiter-family comet 67P/Churyumov-Gerasimenko, and
\citet{2019A&A...630A..24G} reported that solid particles could be of sub-millimeter size and fluffy agglomerates smaller than $1\cm$ are found by MIDAS and GIADA. Therefore, the existence of small grains in the coma must be secondary. \cite{1987A&A...187..753V} found the spatial and temporal variation of small dust grains in Halley's comet coma with Vega-1 and suggested that those small grains are secondary. \cite{1990A&A...231..543B} argued that rotational bursting by radiation pressure proposed by \cite{1980AJ.....85.1538S} is ruled out due to long timescale. 

Moreover, polarimetric observations by \cite{2017MNRAS.469S.475R} and \cite{2020Icar..34813768K} reveal the sharp decrease of polarization at the distance of $r \sim 1.5 - 5\times 10^{3} \km$, followed by a gradual increase with wave-like fluctuations further away. Other recent observations by \cite{2017AJ....154..173K} and \cite{2019A&A...629A.121K} also show the decrease of polarization from the nucleus outward with the cometocentric distance for the inner region of $<5\times 10^{3}\km$. \cite{2020Icar..34813768K} performed modeling for comet 2P/Encke and found that the decreasing size of the dust grains with the cometocentric distance can reproduce observed data. \cite{2004AJ....128.3061J} also suggested that the dis-aggregation of porous grains is required to explain the decrease of polarization with the distance. The author further suggested that spin-up by anisotropic gas flow is a mechanism to disrupt grains into smaller ones of $a<0.1\mum$, although no detailed calculations are presented. The reproduction of small grains from large ones by the RATD mechanism appears to be a plausible mechanism to explain this observational property.



\subsection{Implications for desorption water ice from cometary comae}
Understanding the desorption mechanism (when and where) of water ice from comets is crucial for accurate determination of the nuclei radius and the volume based on the measurement of water vapor. Future space telescopes with SPHEREx \citep{2018arXiv180505489D}, LSST, and JWST \citep{JWST:2018uw} would provide unprecedented data of water ice and vapor, and an accurate determination of ice requires an accurate understanding of their phase transition. The current model of ice vaporization is based on thermal sublimation from icy grains \citep{1979M&P....21..155C}.

Here, we found that water ice mantles could be desorbed from the grain core by radiative torques. The rotational desorption can occur at heliocentric distances of $R_{\rm des}\sim 20$ AU (see Figure \ref{fig:rdes_min}), much larger than the water sublimation radius, $R_{\rm sub}(\H_{2}\O)\sim 3\AU$. Thus, if ice grains could be lifted off the nucleus, then they would be rotationally desorbed. We also quantified the fraction of ice grains desorbed by RATD and find that ice grains that are aligned at high-J attractors could be completely desorbed (see Figure \ref{fig:fdes1}). Measurements of water vapor production rate at $R>R_{\rm sub}((\H_{2}\O)$ could provide a test for the rotational desorption mechanism and constrain the value of $f_{\rm high-J}$, which provide insight into grain geometry and magnetic properties (\citealt{2016ApJ...831..159H}; \citealt{Herranen:2020im}).

\subsection{Implications for Activity of Centaurs and Distant Comets}
Centaurs are icy objects outside Jupiter with a perihelion of $R>5.2\AU$ \citep{1998ApJ...499L.103J}. Some active centaurs are reported in \cite{1998ApJ...499L.103J}, \cite{2009AJ....137.4313J}. Among 23 centaurs observed by \cite{2009AJ....137.4296J}, nine with $R=5.9-8.7 \AU$ are active. Given their far distance from the Sun, their activity is a puzzle because water ice cannot sublimate, whereas CO and CO$_2$ are too volatile, which can drive activity at a larger distance. The possibility that the activity triggered by a mechanism different from thermal sublimation of water ice is suggested. The authors proposed that the activity is triggered by the conversion of amorphous ice into the crystalline form accompanied by the release of trapped gas.

Observations by \cite{2009Icar..201..719M} reveal activity of comets at large heliocentric distances of $R=5.8, 14\AU$, beyond the water sublimation zone of $R_{\rm sub}(\H_{2}\O)\sim 3\AU$ (\citealt{1981Icar...47..342H}; \citealt{2017PASP..129c1001W}). The annealing of amorphous ice to crystalline ice during which gas is released is referred to explain the activity of the comet.

Our calculations show that for low $Q_{\rm gas}$, rotational desorption of ice mantles can occur at $R\sim 20$ AU because of inefficient rotational damping. Note that we assume the presence of water ice grains in the coma and do not discuss the lifting mechanism. However, the low $Q_{\rm gas}$ is consistent with the outgassing of highly volatile ices such as CO and CO$_{2}$. Thus, the rotational desorption could trigger active comets at large distances. Therefore, observations of water vapor at large distances could be possible but require high sensitivity because the amount of water is low due to a small amount of $Q_{\rm gas}$. 

Finally, we find that the amount of water ice desorbed as a function of the heliocentric distance depends on the fraction of grains with high-J, $f_{\rm high-J}$, which depends on the grain shape and magnetic properties (\citealt{2016ApJ...831..159H}; \citealt{2019ApJ...883..122L}). Thus, one can constrain the ice grain properties with observational measurements of the water production rate.

\section{Summary}\label{sec:summary}
We study evolution of dust and water ice grains in comets by radiative torques of sunlight. Our main results are summarized as follows:

\begin{enumerate}

\item We find that RATD can destroy large dust grains of composite structures in cometary comae, resulting in an increase (decrease) in the abundance of small (large) grains with increasing cometocentric distance due to the decrease of the gas density.

\item We suggest the rotational disruption of large grains ($a > 10\mum$) into small fragments ($a \lesssim 0.1\mum$) can explain the time-variability of dust properties observed toward cometary comae, as well as the presence of small dust grains in cometary comae.

\item We study rotational desorption of ice grains and find that water ice grains could be rotationally desorbed rapidly due to centrifugal stress of radiative torques. The rotational desorption could occur at large heliocentric distances much larger than the sublimation radius, $R_{\rm sub}(\H_{2}\O)\sim 3\AU$. Thus, water vapor could be observed from comae at larger distances, albeit it requires high sensitivity due to a low amount of mass released from comet nuclei at large distances. 

\item We suggest the activity of distant comets could be triggered by rotational disruption of dust and ice grains, provided that such grains are lifted off the nucleus by outgassing of highly volatile compounds.

\end{enumerate}

\acknowledgements
We are grateful to the anonymous referee for helpful comments that improved our manuscript. This work was supported by the National Research Foundation of Korea (NRF) grants funded by the Korea government (MSIT) through the Basic Science Research Program (2017R1D1A1B03035359) and Mid-career Research Program (2019R1A2C1087045).


\end{document}